\documentclass[pra,twocolumn,showpacs]{revtex4}
\usepackage{amsmath}
\usepackage{makeidx}
\usepackage{amssymb}
\usepackage{graphicx}
\usepackage[usenames]{color}

\begin{document}

\title{Spontaneous symmetry breaking of Bose-Fermi mixtures \\
in double-well potentials}
\author{S. K. Adhikari$^{1}$\thanks{%
email: adhikari@ift.unesp.br}, B.A. Malomed$^{2}$\thanks{%
email: malomed@post.tau.ac.il}, L. Salasnich$^{3}$\thanks{%
email: salasnich@pd.infn.it}, and F. Toigo$^{3}$\thanks{%
email: flavio.toigo@pd.infn.it}}
\affiliation{$^1$Instituto de F\'isica Te\'orica, 
UNESP $-$ Universidade Estadual
Paulista,01.140-070 S\~ao Paulo, S\~ao Paulo, Brazil \\
$^{2}$Department of Physical Electronics, School of Electrical Engineering,
Faculty of Engineering, Tel Aviv University, Tel Aviv 69978, Israel \\
$^3$CNR and CNISM, Unit\`a di Padova, Dipartimento di Fisica ``Galileo
Galilei'', Universit\`a di Padova, Via Marzolo 8, 35131 Padova, Italy}

\begin{abstract}
We study the spontaneous symmetry breaking (SSB) of a superfluid Bose-Fermi
(BF) mixture in a double-well potential (DWP). The mixture is described by
the Gross-Pitaevskii equation (GPE) for the bosons, coupled to an equation
for the order parameter of the Fermi superfluid, which is derived from the
respective density functional in the unitarity limit (a similar model
applies to the BCS regime too). Straightforward SSB in the degenerate Fermi
gas loaded into a DWP is impossible, as it requires an attractive
self-interaction, while the intrinsic nonlinearity in the Fermi gas is
repulsive. Nonetheless, we demonstrate that the symmetry breaking is
possible in the mixture with attraction between fermions and bosons, like $%
^{40}$\textrm{K} and $^{87}$\textrm{Rb}. Numerical results are
represented by dependencies of asymmetry parameters for both
components on particle numbers of the mixture, $N_{\mathrm{F}}$ and
$N_{\mathrm{B}}$, and by phase diagrams in the $\left(
N_{\mathrm{F}},N_{\mathrm{B}}\right) $ plane, which displays regions
of symmetric and asymmetric ground states. The dynamical picture of
the SSB, induced by a gradual transformation of the single-well
potential into the DWP, is reported too. An analytical approximation
is proposed for the case when GPE for the boson wave function may be
treated by means of the Thomas-Fermi (TF) approximation. Under a
special linear relation between $N_{\mathrm{F}}$ and
$N_{\mathrm{B}}$, the TF approximation allows us to reduce the model
to a single equation for the fermionic function, which includes
competing repulsive and attractive nonlinear terms. The latter one
directly displays the mechanism of the generation of the effective
attraction in the Fermi superfluid, mediated by the bosonic
component of the mixture.
\end{abstract}

\pacs{03.75.Ss, 03.75.Hh, 64.75.+g}
\maketitle

\section{Introduction}

The achievement of quantum degeneracy in bosonic \cite{bose-deg} and
fermionic \cite{fermi-deg} gases of alkali {atoms} has opened the way to the
investigation and manipulation of novel states of atomic matter, such as
Bose-Einstein condensates (BEC) \cite{review,rev-stringa1} and superfluid
Fermi gases \cite{rev-stringa2}. A simple but reliable theoretical tool for
the study of these trapped degenerate gases is the density-functional theory
\cite{lipparini}. In particular, the Gross-Pitaevskii\ equation (GPE), which
accurately describes BECs in dilute gases, is the Euler-Lagrange equation
produced by the Thomas-Fermi (TF) density functional which takes into regard
the inhomogeneity of the condensate \cite{rev-stringa1}. In parallel to
that, many properties of superfluid Fermi gases with balanced (equally
populated) spin components, and the formation of various patterns in them,
can be accurately described, under the conditions of the BCS-BEC crossover,
by an extended TF density functional and its time-dependent version, as it
has been shown recently \cite{manini05,etf,recent,GS}.

One of the fundamental effects in nonlinear media, including BEC, which has
been studied in detail, is the spontaneous symmetry breaking (SSB) in
double-well potentials (DWPs). \emph{Asymmetric} states trapped in symmetric
DWPs are generated by symmetry-breaking bifurcations from obvious symmetric
or antisymmetric states, in the media with the attractive or repulsive
intrinsic nonlinearity, respectively \cite{bifurcation} [the SSB under the
action of competing attractive (cubic) and repulsive (quintic) terms was
studied too \cite{CQ}]. In terms of BEC and other macroscopic quantum
systems, the SSB may be realized as a \textit{quantum phase transition},
which replaces the original symmetric ground state by a new asymmetric one,
when the strength of the self-attractive nonlinearity exceeds a certain
critical value. Actually, a transition of this type was predicted earlier in
the classical context, \textit{viz}., in a model of dual-core nonlinear
optical fibers with the self-focusing Kerr nonlinearity \cite{dual-core}.
Still earlier, the SSB of nonlinear states was studied, in an abstract form,
in the context of the nonlinear Schr\"{o}dinger equation (NLSE) with a
potential term \cite{old}, as well in the discrete self-trapping model \cite%
{Scott}. The latter approach to the description of the SSB effects was later
developed in many works in the form of the two-mode expansion, each mode
representing a mode which is trapped in one of the potential wells (see
Refs. \cite{two-mode} and references therein). As concerns the
interpretation of the SSB as the phase transition, it may be categorized as
belonging to the first or second kind (alias sub- or supercritical SSB
mode), depending on the form of the nonlinearity, spatial dimension, and the
presence or absence of a periodic external potential (an optical lattice)
acting along the additional spatial dimension (if any) \cite{Warsaw}.\ In
the experiment, the self-trapping of asymmetric states has been demonstrated
in the condensate of $^{87}$Rb atoms with repulsive interactions \cite%
{Heidelberg}.

Theoretical studies of the SSB in BECs were extended in various directions.
In particular, the symmetry breaking of matter-wave solitons was predicted
in various two-dimensional (2D) DWP settings \cite{Warsaw}, including the
spontaneous breaking of the \textit{skew symmetry} of solitons and solitary
vortices trapped in double-layer condensates with mutually perpendicular
orientations of quasi-one-dimensional optical lattices induced in the two
layers \cite{skew}. A different variety of the 2D geometry, which gives rise
to a specific mode of the SSB, is based on a symmetric set of four potential
wells \cite{4wells} (a three-well system was considered too \cite{triple}).
Recently, self-trapping of asymmetric states was predicted in the model of
the BEC of dipolar atoms, which interact via long-range forces \cite{DD}.
SSB was also studied in the context of the NLSE with a general nonlinearity
\cite{higher-order}. The symmetry breaking is possible not only in linear
potentials composed of two wells, but also in a similarly structured \textit{%
pseudopotential}, which is produced by a symmetric spatial modulation of the
non-linearity coefficient, with two sharp maxima \cite{Dong}.

Another generalization is the study of the SSB in two- \cite{2comp} and
three-component (spinor) \cite{3comp} BEC mixtures, where the asymmetry of
the density profiles in the two wells is coupled to a difference in
distributions of the different species. Further, the analysis was extended
to a Bose-Fermi (BF) mixture in Ref. \cite{Kivshar}, where a ``frozen"
fermion component was treated as a source of an additional potential for
bosons. Dynamical manifestations of the symmetry breaking (Josephson
oscillations) in a Fermi superfluid trapped in the DWP were recently
considered too \cite{SKA-Pu}.

In spite of many realizations of the SSB studied in the models of degenerate
quantum gases, the self-trapping of stationary asymmetric states has not yet
been considered in fermionic systems. An obvious problem is that a Fermi
gas, loaded into a DWP, cannot feature a direct self-attractive
nonlinearity, which is necessary to induce the SSB in symmetric states.

The objective of the present work is to introduce a model in which the SSB
in a trapped Fermi superfluid is possible due to an effective attraction
\emph{mediated by a bosonic component}, mixed with the fermionic one.
Actually, we consider the SSB in \textit{semi-trapped} BF mixtures, with the
DWP acting on a single species, either the fermionic or bosonic one, as this
setting may be sufficient to hold the entire mixture in the trapping
potential, and induce the SSB in its fermionic component. The analysis will
be performed in the framework of a mean-field model, which couples, via
nonlinear attraction terms, the GPE for the bosonic component\ to an
equation for the fermionic order parameter, derived from the respective
density functional. Inducing an effective boson-mediated attraction between
the fermions requires an attractive {BF} interaction. For this purpose, we
take well-known physical parameters corresponding to the $^{87}\mathrm{Rb}%
-^{40}\mathrm{K}$ mixture, which features repulsion between rubidium atoms
and attraction between the rubidium and potassium, characterized by the
respective positive and negative scattering lengths, $a_{\mathrm{B}}>0$ and $%
a_{\mathrm{BF}}<0$ \cite{modugno}. We consider the case when the
spin-balanced fermionic component of the mixture is in the unitary regime,
corresponding to a diverging scattering length which accounts for the
interaction between the fermionic atoms with opposite orientations of the
spin, $a_{\mathrm{F}}\rightarrow \pm \infty $ (while the BCS regime
corresponds to the vanishingly weak attraction, with $a_{\mathrm{F}%
}\rightarrow -0;$ in either limit, the effective fermionic Lagrangian does
not depend explicitly on $a_{\mathrm{F}}$). In fact, the same model with a
different coefficient of the effective self-repulsion in the Fermi
superfluid applies to the description of the BF mixture with the fermionic
component falling into the BCS regime. Although the self-interaction,
induced by the quantum pressure, in the equation for the fermionic order
parameter is always repulsive, we demonstrate that the SSB in the fermionic
component is indeed possible in the $^{87}\mathrm{Rb}-^{40}\mathrm{K}$
mixture, due to the BF attraction which, as said above, mediates an
effective attraction force in the Fermi superfluid. We also conclude that
the attraction can induce symmetry-preserving or symmetry-breaking
localization of both components in the semi-trapped mixture, depending on
the numbers of the bosons and fermions in it.

The paper is organized as follows. The model is formulated, in a
sufficiently detailed form, in Section II. Results produced by the numerical
analysis are reported in section III, for two variants of the model, with
the DWP acting either only on the fermions, or on the bosons. In Section IV,
we report approximate analytical results, obtained by means of the TF
approximation applied to the GPE for the bosonic wave function. In
particular, assuming a specific linear relation between the fermion and
boson numbers, we can reduce the model to a single equation for the
fermionic wave function with competing self-repulsive and self-attractive
terms, the latter one explicitly demonstrating the mechanism of the
effective attraction between the fermions mediated by ``enslaved" bosons.
The analytical results offer a qualitative explanation to general findings
produced by the numerical analysis. The paper is concluded by Section V.

\section{The model}

Our starting point is a model for the degenerate rarefied quantum gas
composed of $N_{\mathrm{B}}$ condensed bosons of mass $m_{\mathrm{B}}$ and $%
N_{\mathrm{F}}$ fermions of mass $m_{\mathrm{F}}$, in two equally-populated
spin components, at zero temperature. The fermionic component is assumed to
be in the superfluid state at unitarity or, alternatively, in the BCS
regime. The system is made effectively one-dimensional (1D), assuming that
the gas is confined in transverse directions by a tight axisymmetric
harmonic potential, with trapping frequencies $\omega _{\mathrm{\bot B}}$, $%
\omega _{\mathrm{\perp F}}$ for the bosons and fermions, respectively.

Within the framework of the density-functional theory for superfluids \cite%
{etf,recent}, the 3D action of the BF mixture is
\begin{equation}
S=\int \left( \mathcal{L}_{\mathrm{B}}+\mathcal{L}_{\mathrm{F}}+\mathcal{L}_{%
\mathrm{BF}}\right) \ d^{3}\mathbf{r}\ dt\;,  \label{ee}
\end{equation}%
where $\mathcal{L}_{\mathrm{B}}$ is the ordinary bosonic Lagrangian density,
\begin{eqnarray}
\mathcal{L}_{\mathrm{B}} &=&{\frac{i}{2}\hbar \left[ \psi _{\mathrm{B}%
}^{\ast }{\frac{\partial \psi _{\mathrm{B}}}{\partial t}}-\psi _{\mathrm{B}}{%
\frac{\partial \psi _{\mathrm{B}}^{\ast }}{\partial t}}\right] }-{\frac{%
\hbar ^{2}}{2m_{\mathrm{B}}}}\left\vert \nabla \psi _{\mathrm{B}}\right\vert
^{2}  \notag \\
&-&U_{\mathrm{B}}|\psi _{\mathrm{B}}|^{2}-{\frac{2\pi \hbar ^{2}a_{\mathrm{B}%
}}{m_{\mathrm{B}}}}|\psi _{\mathrm{B}}|^{4}\;,  \label{ee-b}
\end{eqnarray}%
$\psi _{\mathrm{B}}(\mathbf{r},t)$ is the macroscopic BEC\ wave function,
and the confining potential for the bosons is
\begin{equation}
U_{\mathrm{B}}(\mathbf{r})={\frac{1}{2}}m_{\mathrm{B}}\omega _{\mathrm{\bot B%
}}^{2}R^{2}+V_{\mathrm{B}}(z)\;,  \label{UB}
\end{equation}%
with $R$ the transverse cylindric radial coordinate and $V_{\mathrm{B}}(z)$
the potential acting in the axial direction $z$. The bosonic superfluid
velocity is $\mathbf{v}_{\mathrm{B}}(\mathbf{r},t)=(\hbar /m_{\mathrm{B}%
})\nabla \theta _{\mathrm{B}}(\mathbf{r},t)$, where $\theta _{\mathrm{B}}(%
\mathbf{r},t)$ is the phase of wave function, $\psi _{\mathrm{B}}(\mathbf{r}%
,t)\equiv \sqrt{n_{\mathrm{B}}(\mathbf{r},t)}e^{i\theta _{\mathrm{B}}(%
\mathbf{r},t)}$, and $n_{\mathrm{B}}(\mathbf{r},t)$ is the bosonic density.

The Galilean-invariant Lagrangian density $\mathcal{L}_{\mathrm{F}}$ of the
Fermi gas with two equally-populated spin components is \cite{etf,sala-new}
\begin{eqnarray}
\mathcal{L}_{\mathrm{F}} &=&i{\frac{\hbar }{4}}{\left[ \psi _{\mathrm{F}}{%
\frac{\partial \psi _{\mathrm{F}}}{\partial t}}-\psi _{\mathrm{F}}{\frac{%
\partial \psi _{\mathrm{F}}}{\partial t}}\right] }-{{\frac{\hbar ^{2}}{8m_{%
\mathrm{F}}}}}|\nabla \psi _{\mathrm{F}}|^{2}  \notag \\
&-&{\frac{3}{5}}\xi {\frac{\hbar ^{2}}{2m_{\mathrm{F}}}}(3\pi
^{2})^{2/3}|\psi _{\mathrm{F}}|^{10/3}-U_{\mathrm{F}}|\psi _{\mathrm{F}%
}|^{2}\;,  \label{ee-f}
\end{eqnarray}%
where $\psi _{\mathrm{F}}(\mathbf{r},t)$ is the superfluid order parameter
of the Fermi gas at unitary \cite{etf}, $2m_{\mathrm{F}}$ is the mass of {a
pair of fermions with spins up and down}, and the potential acting on the
fermionic atoms is
\begin{equation}
U_{\mathrm{F}}(\mathbf{r})={\frac{1}{2}}m_{\mathrm{F}}\omega _{\mathrm{\bot F%
}}^{2}R^{2}+V_{\mathrm{F}}(z),
\end{equation}%
cf. its bosonic counterpart (\ref{UB}). The fermionic superfluid velocity is
$\mathbf{v}_{\mathrm{F}}(\mathbf{r},t)=(\hbar /2m_{\mathrm{F}})\nabla \theta
_{\mathrm{F}}(\mathbf{r},t)$, where $\theta _{\mathrm{F}}(\mathbf{r},t)$ is
the phase of the order parameter, $\psi _{\mathrm{F}}(\mathbf{r},t)\equiv
\sqrt{n_{\mathrm{F}}(\mathbf{r},t)}e^{i\theta _{\mathrm{F}}(\mathbf{r},t)}$,
and $n_{\mathrm{F}}(\mathbf{r},t)$ is the density of fermionic atoms.
Constant $\xi $ in expression (\ref{ee-f}) is $\xi =1$ in the deep BCS
regime, and $\xi \simeq 0.4$ at the unitarity \cite{rev-stringa2}. In
calculations reported below, we fixed $\xi =0.45$, assuming the unitarity
regime. Lastly, the Lagrangian density $\mathcal{L}_{\mathrm{BF}}$ in
expression (\ref{ee}) accounting for the BF interaction is
\begin{equation}
\mathcal{L}_{\mathrm{BF}}=-{\frac{2\pi \hbar ^{2}a_{\mathrm{BF}}}{m_{R}}}%
\,|\psi _{\mathrm{B}}|^{2}\ |\psi _{\mathrm{F}}|^{2}\;,  \label{ee-bf}
\end{equation}%
where $m_{\mathrm{R}}=m_{\mathrm{B}}m_{\mathrm{F}}/(m_{\mathrm{B}}+m_{%
\mathrm{F}})$ is the respective reduced mass, and as said above, $a_{\mathrm{%
BF}}<0$ corresponds to an attractive BF interaction, which is necessary to
support the SSB of the fermionic component in the presence of the DWP. We
stress our assumptions: i) $a_{\mathrm{BF}}$ is independent of the spin
component; ii) the fermionic density profiles are identical for the two
components.

In the nearly 1D configuration, transverse widths of the atomic
distributions are determined by the width of the ground states of the
respective harmonic oscillators:
\begin{equation}
a_{\mathrm{\bot B}}=\sqrt{\hbar /(m_{\mathrm{B}}\omega _{\mathrm{\bot B}})}%
,a_{\mathrm{\bot F}}=\sqrt{\hbar /(2m_{\mathrm{F}}\omega _{\mathrm{\bot F}})}%
,  \label{perp}
\end{equation}%
for the condensed bosons and superfluid fermions at the unitarity.
The boson and fermion components exhibit the effective 1D behavior
if their chemical potentials are much smaller than the corresponding
transverse-trapping energies, $\hbar \omega _{\mathrm{\bot B}}$ and
$\hbar \omega _{\mathrm{\bot F}}/2$. Notice that the effective mass
$2m_{F}$ in the fermionic harmonic length of Eq. (\ref{perp}) is a
direct consequence of the coefficient $1/8$ in gradient term
$|\nabla \psi _{F}|^{2}$ of the fermionic Lagrangian (\ref{ee-f}).
This coefficient is chosen to reproduce very accurately the Monte
Carlo simulations of the 3D unitary Fermi gas
\cite{etf,recent,sala-new}.
Under these conditions, we can adopt the known factorized \textit{ans\"{a}tze%
} for the 3D wave functions \cite{Luca},
\begin{eqnarray}
\psi _{\mathrm{B}}(\mathbf{r},t) &=&\sqrt{N_{\mathrm{B}}}\ {\frac{%
e^{-R^{2}/(2a_{\mathrm{\bot B}}^{2})}}{\pi ^{1/2}a_{\mathrm{\bot B}}}}\ \Phi
_{\mathrm{B}}(z,t)\;,  \label{ans1} \\
\psi _{\mathrm{F}}(\mathbf{r},t) &=&\sqrt{N_{\mathrm{F}}}\ {\frac{%
e^{-R^{2}/(2a_{\mathrm{\bot F}}^{2})}}{\pi ^{1/2}a_{\mathrm{\bot F}}}}\ \Phi
_{\mathrm{F}}(z,t)\;,  \label{ans2}
\end{eqnarray}%
where $\Phi _{\mathrm{B}}(z)$ and $\Phi _{\mathrm{F}}(z)$ are the 1D (axial)
wave functions, which are subject to the usual normalization conditions,
\begin{equation}
\int_{-\infty }^{+\infty }|\Phi _{\mathrm{B}}(z,t)|^{2}\ dz=\int_{-\infty
}^{+\infty }|\Phi _{\mathrm{F}}(z,t)|^{2}\ dz=1.  \label{con}
\end{equation}%
Inserting expressions (\ref{ans1}) and (\ref{ans2}) into Eq. (\ref{ee}), the
action can be written as $S=S^{\mathrm{(1D)}}-\left[ N_{\mathrm{B}}\ \hbar
\omega _{\mathrm{\bot B}}-N_{\mathrm{F}}\ \hbar \omega _{\mathrm{\bot F}}/2%
\right] t$, with the effective 1D action
\begin{equation}
S^{\mathrm{(1D)}}=\int dt\int_{-\infty }^{+\infty }dz\left[ \mathcal{L}_{%
\mathrm{B}}^{\mathrm{(1D)}}+\mathcal{L}_{\mathrm{F}}^{\mathrm{(1D)}}+%
\mathcal{L}_{\mathrm{BF}}^{\mathrm{(1D)}}\right] ,  \label{e}
\end{equation}%
where the usual 1D Gross-Pitaevskii Lagrangian density is
\begin{eqnarray}
\mathcal{L}_{B}^{\mathrm{(1D)}} &=&N_{\mathrm{B}}\left( {{\ i\frac{\hbar }{2}%
\left[ \Phi _{\mathrm{B}}^{\ast }{\frac{\partial \Phi _{\mathrm{B}}}{%
\partial t}}-\Phi _{\mathrm{B}}{\frac{\partial \Phi _{\mathrm{B}}^{\ast }}{%
\partial t}}\right] }}-{\frac{\hbar ^{2}}{2m_{\mathrm{B}}}}\left\vert {\frac{%
\partial \Phi _{\mathrm{B}}}{\partial z}}\right\vert ^{2}\right.   \notag \\
&&\left. -V_{\mathrm{B}}|\Phi _{\mathrm{B}}|^{2}-{\frac{1}{2}}G_{\mathrm{B}%
}|\Phi _{\mathrm{B}}|^{4}\right) ,  \label{e-b}
\end{eqnarray}%
where the boson self-interaction strength in 1D is
\begin{equation}
G_{\mathrm{B}}\equiv 2N_{\mathrm{B}}\hbar \omega _{\mathrm{\bot B}}\,a_{%
\mathrm{B}}.  \label{GB}
\end{equation}%
Further, the 1D fermionic Lagrangian density in Eq. (\ref{e}) is
\begin{eqnarray}
\mathcal{L}_{\mathrm{F}}^{\mathrm{(1D)}} &=&N_{\mathrm{F}}\left( i{\frac{%
\hbar }{4}}{\left[ {\Phi _{\mathrm{F}}^{\ast }{\frac{\partial \Phi _{\mathrm{%
F}}}{\partial t}}-\Phi _{\mathrm{F}}{\frac{\partial \Phi _{\mathrm{F}}^{\ast
}}{\partial t}}}\right] }-{\frac{\hbar ^{2}}{8m_{\mathrm{F}}}}\left\vert {%
\frac{\partial \Phi _{\mathrm{F}}}{\partial z}}\right\vert ^{2}\right.
\notag \\
&&\left. -{\frac{3}{5}}A|\Phi _{\mathrm{F}}|^{10/3}-V_{\mathrm{F}}|\Phi _{%
\mathrm{F}}|^{2}\right) ,  \label{e-f}
\end{eqnarray}%
with the effective strength of the fermionic quantum pressure,
\begin{equation}
A=(3\pi ^{2})^{2/3}(3\xi /5)\hbar ^{2}N_{\mathrm{F}}^{2/3}/(2m_{\mathrm{F}%
}a_{\mathrm{\bot F}}^{4/3}),  \label{A}
\end{equation}%
emerging as the coefficient in front of the bulk kinetic energy of the Fermi
gas in the unitarity limit. Finally, the 1D Lagrangian density of the BF
interaction is
\begin{equation}
\mathcal{L}_{\mathrm{BF}}^{\mathrm{(1D)}}=-N_{\mathrm{B}}N_{\mathrm{F}}\ G_{%
\mathrm{BF}}\,|\Phi _{\mathrm{B}}|^{2}\ |\Phi _{\mathrm{F}}|^{2},
\label{e-bf}
\end{equation}%
where the corresponding 1D interaction strength is%
\begin{equation}
G_{\mathrm{BF}}\equiv 2\hbar ^{2}a_{\mathrm{BF}}/(m_{\mathrm{R}}a_{\mathrm{%
\bot B}}a_{\mathrm{\bot F}}).  \label{BF}
\end{equation}

For numerical calculations, we set $a_{\mathrm{\perp B}}=a_{\mathrm{\perp F}%
}\equiv a_{\perp }$ and $\omega _{\mathrm{\perp B}}=\omega _{\perp \mathrm{F}%
}\equiv \omega _{\perp }$, measuring lengths and time in units of $a_{\perp }
$ and $\omega _{\perp }^{-1}$, respectively. This implies that $2m_{\mathrm{F%
}}=m_{\mathrm{B}}$ (hence, $m_{\mathrm{R}}=m_{\mathrm{B}}/3$), a condition
which is roughly satisfied by the $^{87}\mathrm{Rb}-^{40}\mathrm{Kb}$
mixture. We consider $^{40}$K atoms in the two equally-populated hyperfine
states $|F=9/2,m_{F}=-9/2\rangle $ and $|F=9/2,m_{F}=-7/2\rangle $, and the $%
^{87}$Rb atoms in the hyperfine state $|F=2,m_{F}=2\rangle $. This mixture
is a good candidate for experimental study of the SSB because the BF
scattering length is negative, as stressed above, $a_{\mathrm{BF}}\approx
-284a_{0}$, where $a_{0}$ is the Bohr radius \cite{modugno}. Simultaneously,
the scattering length for collisions between rubidium atoms is positive, $a_{%
\mathrm{B}}\approx 108a_{0}$ \cite{review,modugno,Volz}. In principle, the
magnetic field inducing the Feshbach resonance can affect the values of $a_{%
\mathrm{BF}}$ and $a_{\mathrm{B}}$, but we neglect this effect in all the
calculations reported in the present work.

The application of the variational procedure to the effective action (\ref{e}%
) produces a system of coupled NLSEs,
\begin{eqnarray}
&&\left[ -\frac{1}{2}{\frac{\partial ^{2}}{\partial x^{2}}}+W_{\mathrm{B}%
}(x)+g_{\mathrm{B}}|\phi _{\mathrm{B}}|^{2}\right.  \notag \\
&&\left. +g_{\mathrm{BF}}N_{\mathrm{F}}|\phi _{\mathrm{F}}|^{2}\right] \phi
_{\mathrm{B}}=i{\frac{\partial }{\partial \tau }}\phi _{\mathrm{B}}\;,
\label{uB}
\end{eqnarray}%
\begin{eqnarray}
&&\left[ -\frac{1}{8}{\frac{\partial ^{2}}{\partial x^{2}}}+g_{\mathrm{F}%
}|\phi _{\mathrm{F}}|^{4/3}+W_{\mathrm{F}}(x)\right.  \notag \\
&&\left. +g_{\mathrm{BF}}N_{\mathrm{B}}|\phi _{\mathrm{B}}|^{2}\right] \phi
_{\mathrm{F}}={\frac{i}{2}}{\frac{\partial }{\partial \tau }}\phi _{\mathrm{F%
}}\;,  \label{uF}
\end{eqnarray}%
where $x=z/a_{\bot }$, $\tau =\omega _{\bot }t$, and $\phi _{\mathrm{B}%
}=a_{\bot }^{1/2}\Phi _{\mathrm{B}}$, $\phi _{\mathrm{F}}=a_{\bot
}^{1/2}\Phi _{\mathrm{F}}$, $W_{\mathrm{B}}=V_{\mathrm{B}}/(\hbar \omega
_{\bot })$, $W_{\mathrm{F}}=V_{\mathrm{F}}/(\hbar \omega _{\bot })$, and the
renormalized interaction coefficients are derived from expressions (\ref{GB}%
), (\ref{A}), and (\ref{BF}),%
\begin{eqnarray}
g_{\mathrm{B}} &=&G_{\mathrm{B}}/(\hbar \omega _{\bot }a_{\bot })\equiv
2\left( a_{\mathrm{B}}/a_{\perp }\right) N_{\mathrm{B}},  \notag \\
g_{\mathrm{BF}} &=&G_{\mathrm{BF}}/(\hbar \omega _{\bot }a_{\bot })\equiv
6a_{\mathrm{BF}}/a_{\perp },  \notag \\
g_{\mathrm{F}} &=&A/(a_{\bot }^{2/3}\hbar \omega _{\bot })\equiv \left( 3\pi
^{2}\right) ^{2/3}\left( 3\xi /5\right) N_{\mathrm{F}}^{2/3}.  \label{g}
\end{eqnarray}%
Note that the rescaled wave functions are subject to the same normalization
conditions as in Eqs. (\ref{con}), i.e.,
\begin{equation}
\int_{-\infty }^{+\infty }|\phi _{\mathrm{B}}(x,\tau )|^{2}\
dx=\int_{-\infty }^{+\infty }|\phi _{\mathrm{F}}(x,\tau )|^{2}\ dx=1.
\label{con'}
\end{equation}%
If condition $2m_{\mathrm{F}}=m_{\mathrm{B}}$ does not hold, the coupled
equations can be cast in the same form, with a difference that coefficient $%
2m_{\mathrm{F}}/m_{\mathrm{B}}$ appears in front of the second derivative in
Eq. (\ref{uB}).

The coupled NLSEs in the form of Eqs. (\ref{uF}) actually generalize the
static and dynamical equations for BF systems which were used in various
settings in Refs. \cite{BFequations}. In particular, the
semi-phenomenological equations used for the study of 1D gap solitons in the
BF mixture in Ref. \cite{we} are also tantamount to Eqs. (\ref{uF}), up to a
difference in coefficients.

To find stationary solutions to Eqs. ((\ref{uB}) and (\ref{uF}), we employed
a Crank-Nicolson finite-difference scheme for simulations of the equations
in imaginary time, {{using the Fortran codes provided in Ref. \cite%
{sadhan-num}. We employed space and time steps }}$\Delta x={0.025}${\ and }${%
\Delta t=0.001}$,{\ and a sufficiently large number of iterations to ensure
the convergence. The stability of the stationary solutions against small
perturbations was then tested by simulations in real time. }

Due to the attractive character of the BF interactions ($a_{\mathrm{BF}}<0$%
), the true ground state of the 3D mixture collapses towards energy $%
E=-\infty $. Nevertheless, because of the strong transverse confinement, the
quasi-1D metastable state has an indefinitely long lifetime \cite{mix-io}.
Actually, stationary solutions generated by the imaginary-time integration
represent the ground state of the effective one-dimensional BF system based
on Eqs. (\ref{uB}) and (\ref{uF}) -- in the same sense as the famous
matter-wave solitons realize the ground state of the quasi-1D condensate of $%
^{7}\mathrm{Li}$ atoms \cite{solitons}.

\section{Results of the numerical analysis}

\subsection{Axially trapped fermions and free bosons}

We start the analysis by considering the configuration with the DWP acting
solely on the fermionic component:
\begin{equation}
W_{\mathrm{F}}(x)=\alpha _{\mathrm{F}} x^{2}+\beta _{\mathrm{F}}\exp \left(
-\gamma _{\mathrm{F}}x^{2}\right) ,~W_{\mathrm{B}}(z)=0,  \label{pot-f}
\end{equation}
where all constants are positive. An elementary consideration of this
potential demonstrates that it features the double-well structure provided
that $\alpha _{\mathrm{F}}<\beta _{\mathrm{F}}\gamma _{\mathrm{F}}$, with
two symmetric potential minima located at points%
\begin{equation}
x_{\min }=\pm \gamma _{\mathrm{F}}^{-1/2}\sqrt{\ln \left( \beta _{\mathrm{F}%
}\gamma _{\mathrm{F}}/\alpha _{\mathrm{F}}\right) }.  \label{min}
\end{equation}%
We reports results of simulations for $\alpha _{\mathrm{F}}=1/2$, $\beta _{%
\mathrm{F}}=16$, and $\gamma _{\mathrm{F}}=10$, which adequately represents
the generic situation; in this case, Eq. (\ref{min}) yields $x_{\min
}\approx \pm \allowbreak 0.76$.

The purpose of the analysis is to construct $x$-symmetric and asymmetric
ground states of the system, varying the control parameters, and identify
the respective \textit{SSB bifurcation}, i.e., the transition to the
asymmetric ground state. The asymmetry of its Fermi and Bose components is
characterized by parameters
\begin{equation}
\theta _{\mathrm{F,B}}\equiv \frac{\int_{0}^{\infty }|\phi _{\mathrm{F,B}%
}(x)|^{2}dx-\int_{-\infty }^{0}|\phi _{\mathrm{F,B}}(x)|^{2}dx}{%
\int_{-\infty }^{+\infty }|\phi _{\mathrm{F,B}}(x)|^{2}dx}\;.  \label{theta}
\end{equation}%
Recall that the denominator in this expression is actually $1$ for both
species, as per Eq. (\ref{con'}).

\begin{figure}[tbp]
\begin{center}
\includegraphics[width=.65\linewidth]{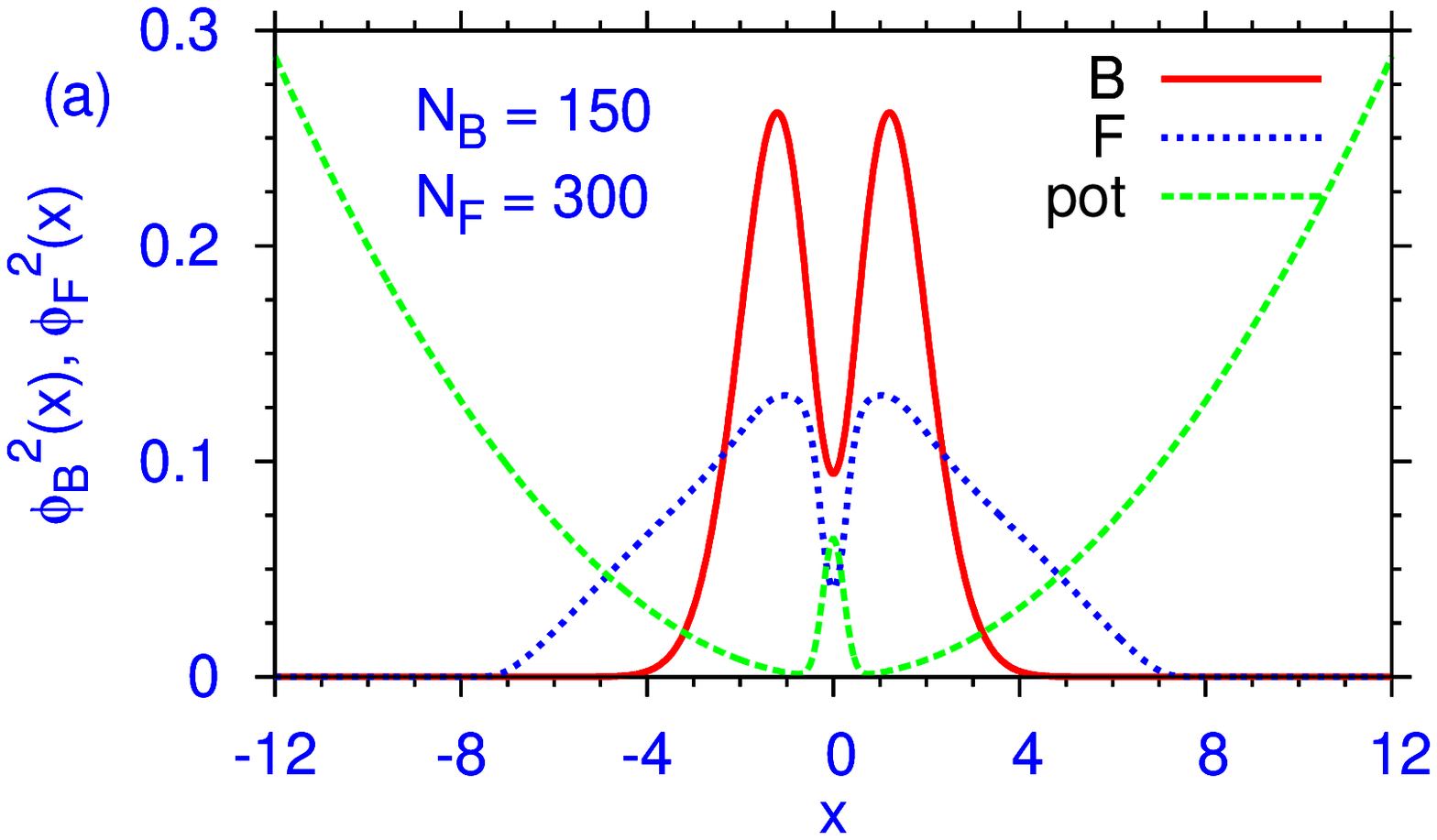} \includegraphics[width=.65%
\linewidth]{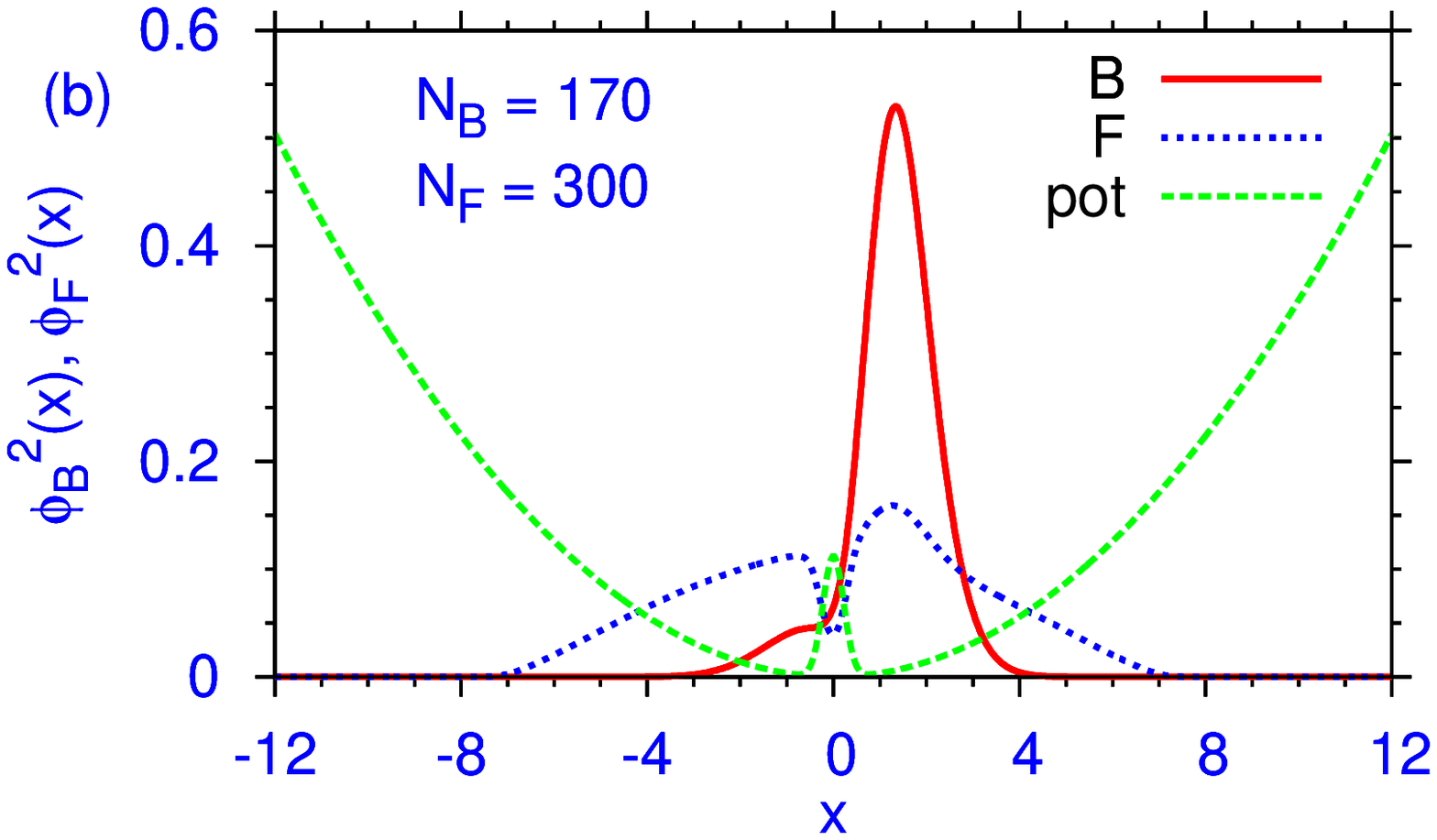} \includegraphics[width=.65\linewidth]{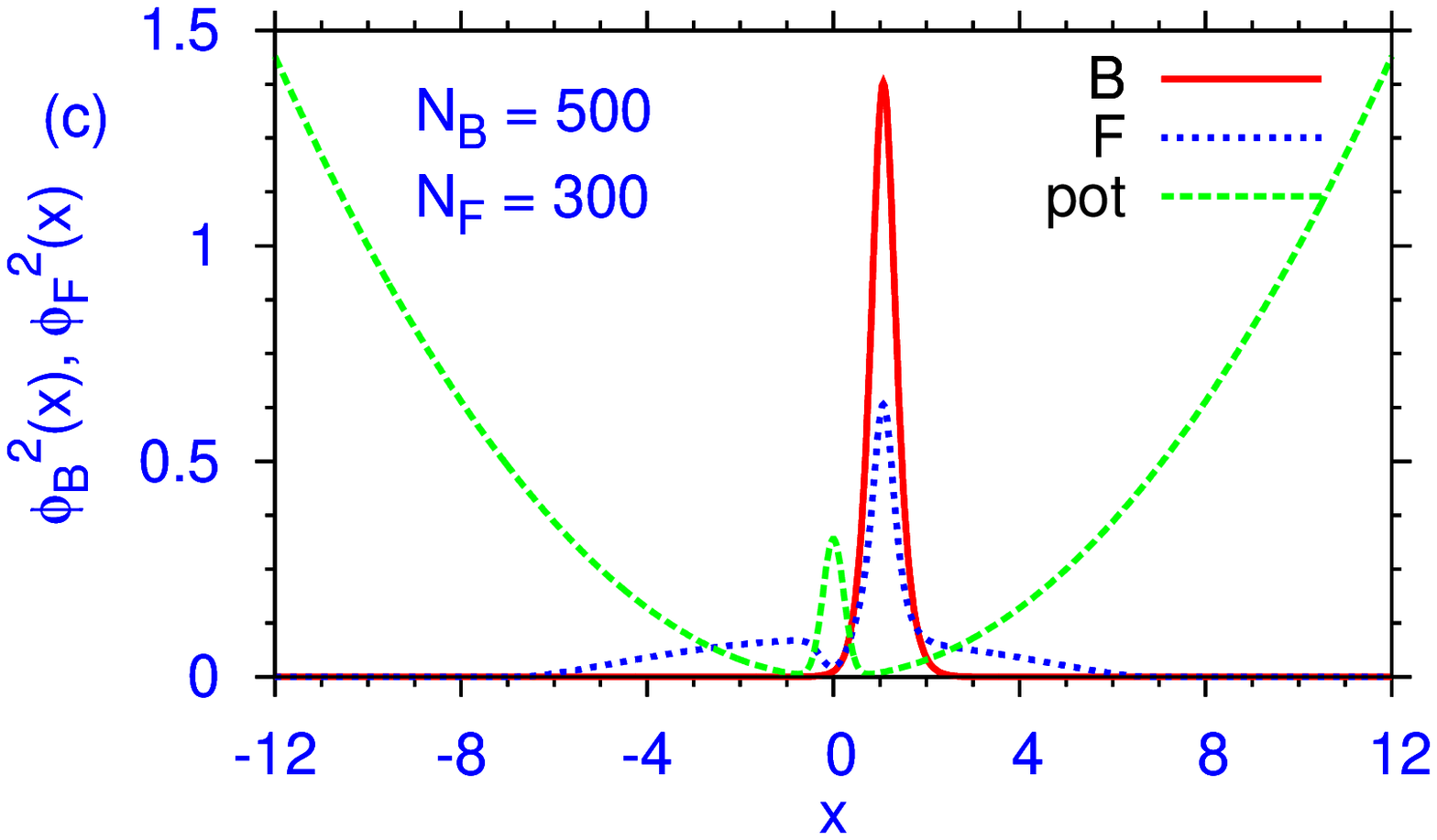}
\end{center}
\caption{(Color online). Density profiles of the $^{87}\mathrm{Rb}-^{40}%
\mathrm{Kb}$ mixture, $\protect\phi _{\mathrm{B}}^{2}(x)$ and $\protect\phi %
_{\mathrm{F}}^{2}(x)$, marked by the respective labels, in the case of the
double-well potential (shown in arbitrary units by curve ``pot") acting on
the fermions, while no axial potential is applied to the bosons. The three
panels differ by the number of bosons, $N_{\mathrm{B}}$, as indicated in
each panel. Recall that the axial coordinate $x$ is measured in units of the
transverse-confinement length, $a_{\bot }$, while $\protect\phi _{\mathrm{B}%
}^{2}$ and $\protect\phi _{\mathrm{F}}^{2}$ are displayed in units of $%
a_{\bot }^{-1}$.}
\label{ssb-f1}
\end{figure}

In Fig. \ref{ssb-f1} we display a set of axial (1D) profiles of the
densities of both components in the ground state, generated by the
integration of Eqs. (\ref{uB}) and (\ref{uF}) in imaginary time, in the case
of the BF attraction and weak \emph{repulsion }between the bosons. The
respective values of the interaction coefficients in Eqs. (\ref{uB}) and (%
\ref{uF}) are taken as per Eqs. (\ref{g}), with the above-mentioned values
of $a_{\mathrm{BF}}$ and $a_{\mathrm{B}}$ for the $^{87}\mathrm{Rb}-^{40}%
\mathrm{Kb}$ mixture and a fixed transverse-trapping length, $a_{\bot }=1\
\mathrm{\mu }$m, substituting various values of atomic numbers $N_{\mathrm{B}%
}$ and $N_{\mathrm{F}}$. The figure clearly shows the transition from the
symmetric ground state to the asymmetric one, with the increase of the
number of bosons, $N_{\mathrm{B}}$, i.e., as a matter of fact, the strength
of both the boson-boson\ and BF interactions. Accordingly, the increase of $%
N_{\mathrm{B}}$ leads to a stronger overlap between the bosons and fermions,
and to the SSB, i.e., the transition from symmetric ground states to an
asymmetric one, which happens (simultaneously in both components) between $%
N_{\mathrm{B}}=150$ and $170$, for a fixed number of fermions, $N_{\mathrm{F}%
}=300$.

\begin{figure}[tbp]
\begin{center}
\includegraphics[width=0.8\linewidth]{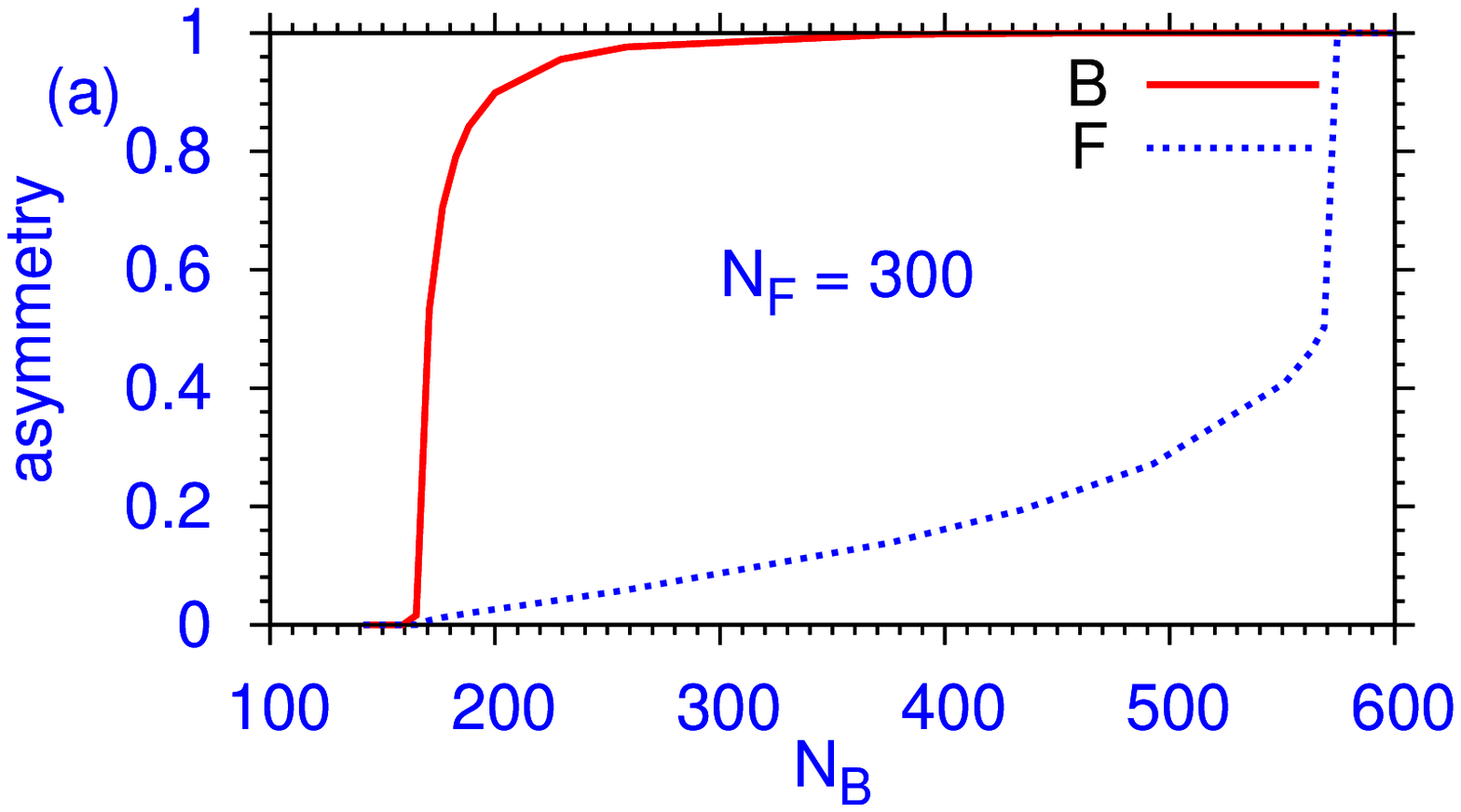} \includegraphics[width=0.8%
\linewidth]{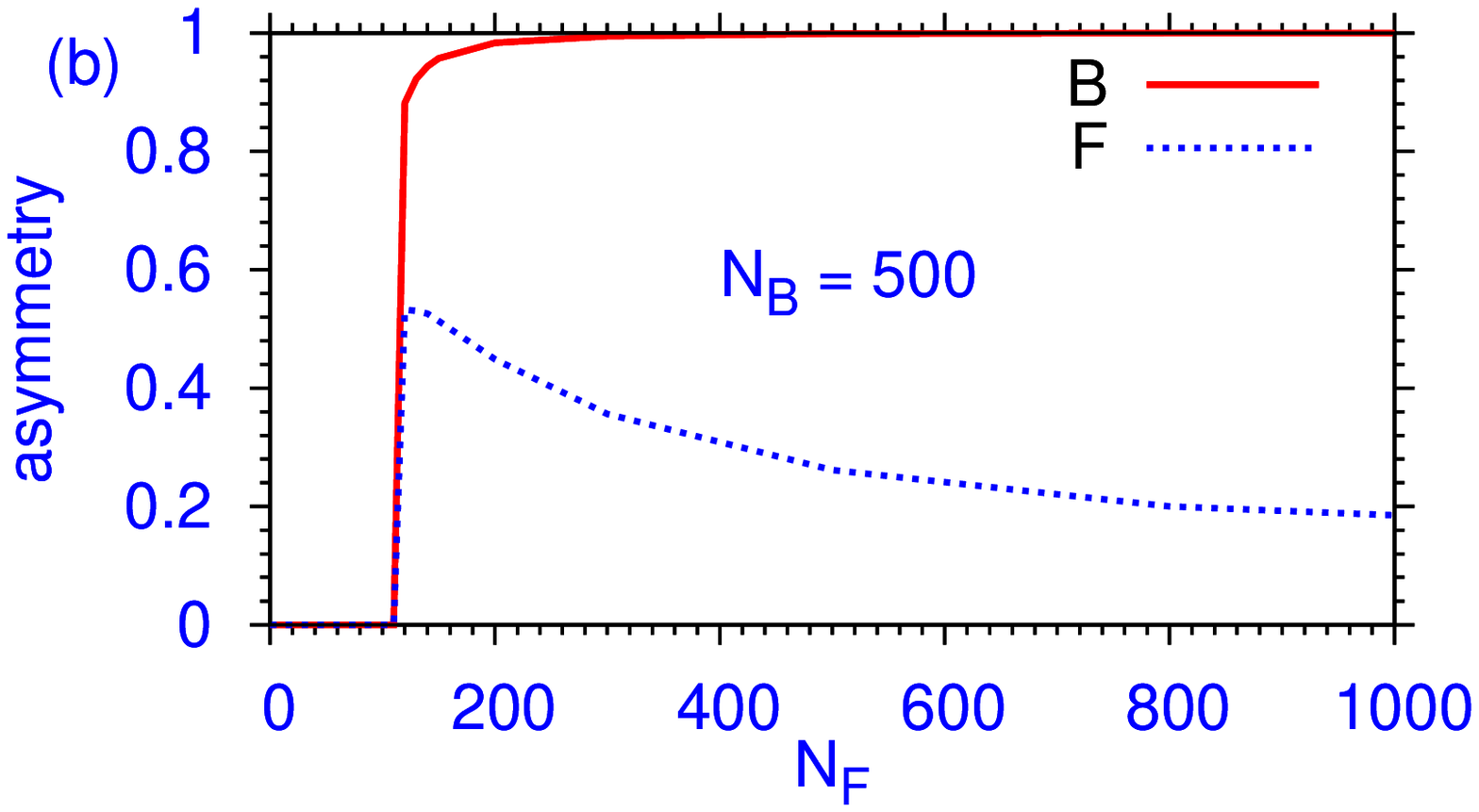}
\end{center}
\caption{(Color online). Asymmetry parameters (\protect\ref{theta}), for the
bosons (B) and fermions (F) in the $^{87}\mathrm{Rb}-^{40}\mathrm{Kb}$
mixture, loaded into potential (\protect\ref{pot-f}), as functions of: (a)
the number of bosons, $N_{\mathrm{B}}$, at a fixed number of fermions, $N_{%
\mathrm{F}}=300$; (b) the number of fermions, at fixed $N_{\mathrm{B}}=500$.}
\label{ssb-f2}
\end{figure}

Similarly to what is shown in Fig. \ref{ssb-f1}, the symmetric ground state
is replaced by an asymmetric one with the increase of $N_{\mathrm{F}}$ at
fixed $N_{\mathrm{B}}$, as this implies the strengthening of the BF
interaction. The summary of the results for the SSB in the present setting
is provided in Fig. \ref{ssb-f2} by plots of asymmetry parameters (\ref%
{theta}) versus $N_{\mathrm{B}}$ for fixed $N_{\mathrm{F}}$, and vice versa.
Naturally, the SSB of both components happens at the same point [for
instance, at $N_{\mathrm{B}}\approx 105$ in panel (a)]. Nevertheless, the
resulting bosonic asymmetry is essentially stronger [in Fig. \ref{ssb-f1}(a)
-- up to a point, $N_{\mathrm{B}}\approx 580$, at which both $\theta _{%
\mathrm{B}}$ and $\theta _{\mathrm{F}}$ attain values very close to $1$,
i.e., practically all the atoms are collected in a single potential well].
In Fig. \ref{ssb-f1}(b), the behavior of the fermionic asymmetry, $\Theta _{%
\mathrm{F}}$, is different: it jumps to a maximum value at the SSB point,
and then gradually decreases. This difference between the bosonic and
fermionic components is natural, as the intrinsic repulsion in the latter
one tends to restore the symmetry between the distributions of atoms in the
two potential wells.

\begin{figure}[tbp]
\begin{center}
\includegraphics[width=0.8\linewidth]{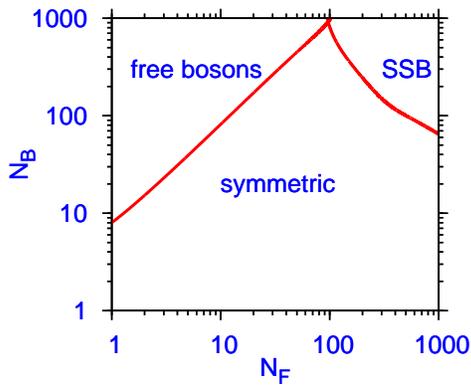}
\end{center}
\caption{(Color online). The phase diagram of the $^{87}\mathrm{Rb}-^{40}%
\mathrm{Kb}$ mixture in potential (\protect\ref{pot-f}), which acts only on
the fermions. The diagram shows, on the logarithmic scales in the $\left( N_{%
\mathrm{B}},N_{\mathrm{F}}\right) $ plane, regions of the symmetric ground
state, and of the spontaneous symmetry breaking (SSB). ``Free bosons"
implies delocalization of the bosonic wave function.}
\label{ssb-f3}
\end{figure}

The results are further summarized in Fig. \ref{ssb-f3}, which displays the
phase diagram of the mixture. There are three regions in the $(N_{\mathrm{F}%
},N_{\mathrm{B}})$ plane: an area where the attraction to fermions cannot
keep bosonic atoms in the trapped state (``free bosons"), the region where
the bosons and fermions are trapped in the symmetric state, with respect to
the DWP, and the SSB region, where the mixture is trapped in the asymmetric
ground state.

An example of the dynamical development of the SSB from an initially
symmetric configuration, in the case where the ground state is asymmetric,
is presented by Fig. \ref{ssb-f3}. Initially, the bosons and fermions form a
stable symmetric bound state, via their mutual attraction, in the \emph{%
single-well} potential acting on the fermions, which is taken in the form of
(\ref{pot-f}) with $\beta _{\mathrm{F}}=0$. Then, $\beta _{\mathrm{F}}$ is
ramped linearly in time ($0<t<80$) from $\beta _{\mathrm{F}}=0$ to $\beta _{%
\mathrm{F}}=16$, which leads to splitting the single potential well into
two, as per Eq. (\ref{min}). The dynamical picture clearly shows the
transition of the initial symmetric state into the symmetry-broken one. Both
components get spontaneously collected in one of the wells, where they stay
together due to the mutual attraction, approaching an equilibrium
configuration.

\begin{figure}[tbp]
\begin{center}
\includegraphics[width=.49\linewidth]{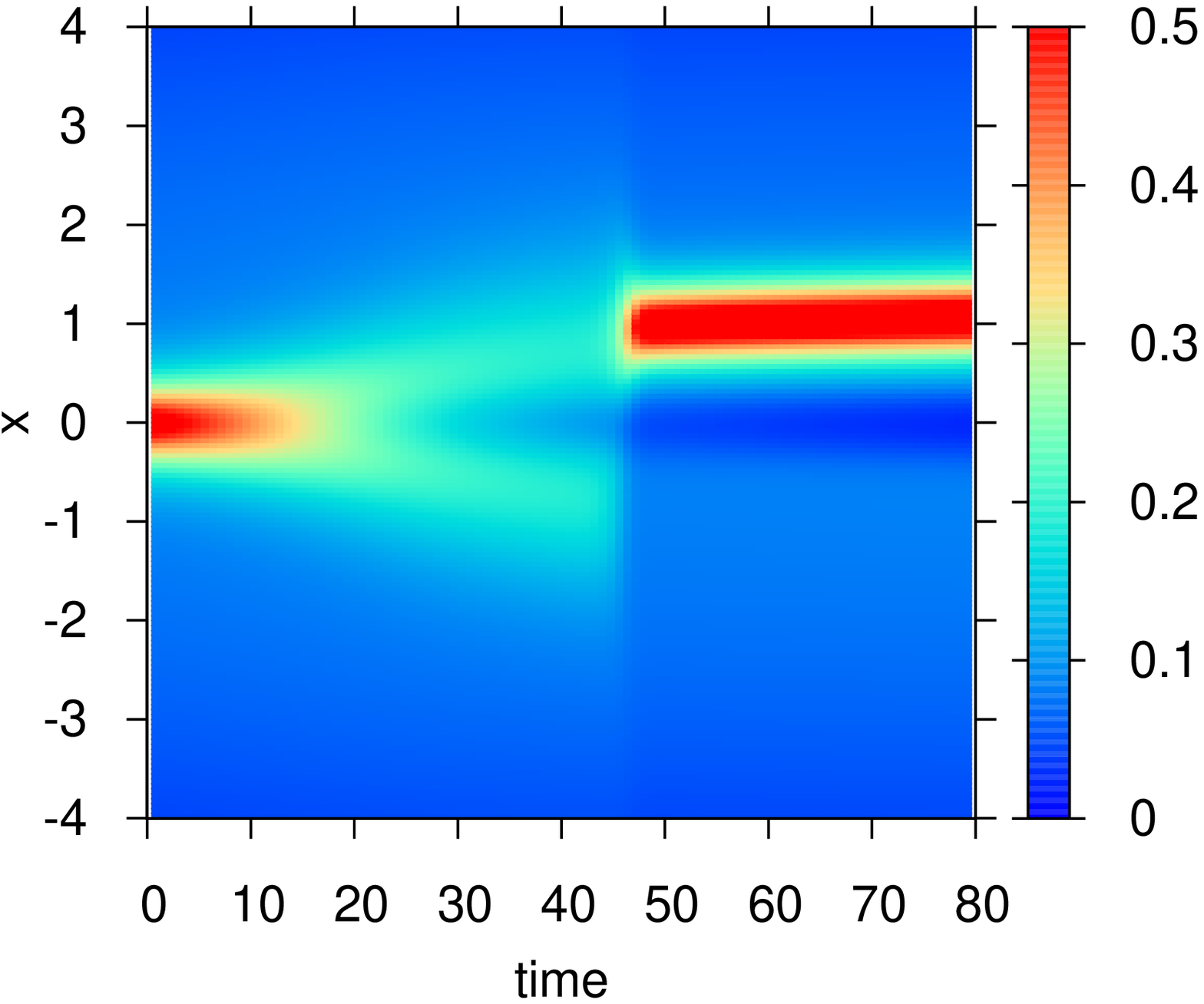} \includegraphics[width=.49%
\linewidth]{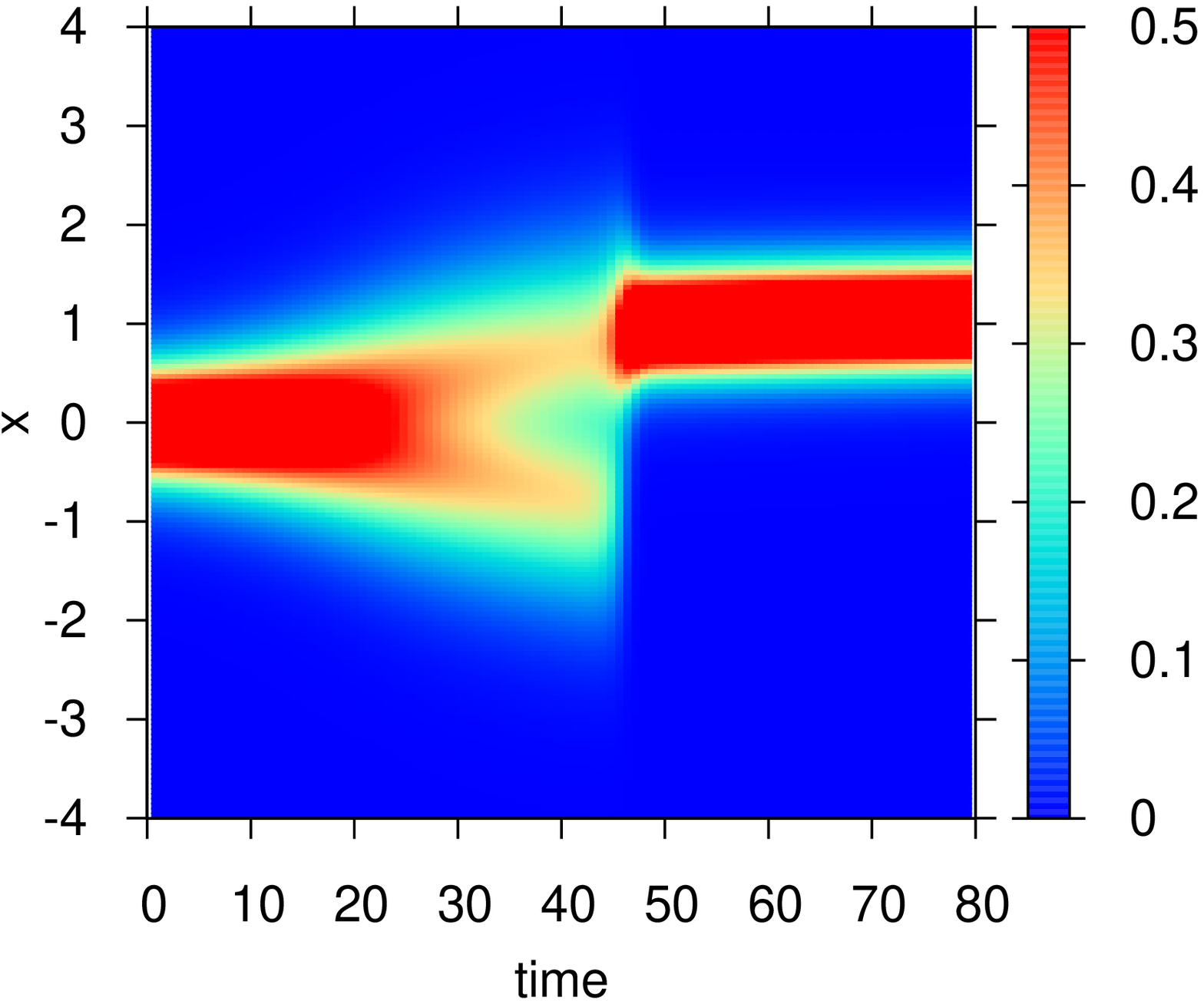}
\end{center}
\caption{(Color online). Contour plots for the evolution of densities $|%
\protect\phi _{\mathrm{B}}(x,t)|^{2}$ and $|\protect\phi _{\mathrm{F}%
}(x,t)|^{2}$ of the (a) bosons and (b) fermions in the $^{87}\mathrm{Rb}%
-^{40}\mathrm{Kb}$ mixture. At $t=0$, the fermions are trapped in the
\textit{single-well} potential (\protect\ref{pot-f}), with $\protect\beta _{%
\mathrm{F}}=0$. Then, $\protect\beta _{\mathrm{F}}$ linearly increases from $%
0$ to $16$ by $t=80$, which implies the transition to the double-well
potential with well-separated symmetric minima. Numbers of particles are $N_{%
\mathrm{B}}=500$ and $N_{\mathrm{F}}=300$.}
\label{ssb-f4}
\end{figure}

\subsection{Axially trapped bosons and free fermions}

Now, we consider the action of the DWP on the bosons only, taking the
potential as
\begin{equation}
W_{\mathrm{B}}(x)=\alpha _{\mathrm{B}}x^{2}+\beta _{\mathrm{B}}\exp \left(
-\gamma _{\mathrm{B}}x^{2}\right) ,~W_{\mathrm{F}}=0  \label{pot-b}
\end{equation}%
For this setting, numerical results are presented with $\alpha _{\mathrm{B}%
}=1/2$, $\beta _{\mathrm{B}}=16$, and $\gamma _{\mathrm{B}}=10$, i.e., the
same parameters of the DWP as used above for the trapping the fermions. We
again aim to construct the ground state of the system as a function of the
control parameters, and investigate its spontaneous transition into an
asymmetric shape.

\begin{figure}[tbp]
\begin{center}
\includegraphics[width=0.8\linewidth]{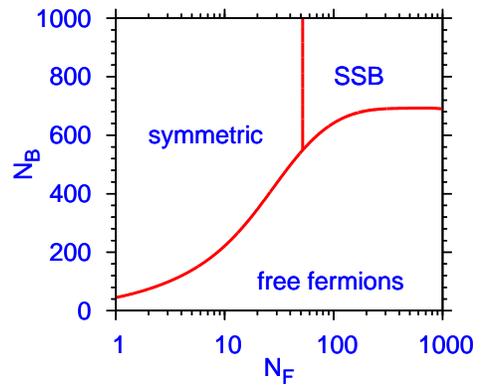}
\end{center}
\caption{(Color online). The phase diagram of the $^{87}\mathrm{Rb}-^{40}%
\mathrm{K}$ mixture loaded into potential (\protect\ref{pot-b}) which acts
only on the bosons.}
\label{ssb-f5}
\end{figure}

Results of the analysis for this setting are summarized in the respective
phase diagram of the mixture plotted in Fig. \ref{ssb-f5} (cf. Fig. \ref%
{ssb-f3}). In this case too, there are three regions in the $(N_{\mathrm{F}%
},N_{\mathrm{B}})$ plane: an area where the fermions cannot be held in a
localized state by the attraction to bosons (``free fermions"), the region
where the fermions are trapped, along with the bosons, in a symmetric ground
state, and the region where the trapped ground state is asymmetric (``SSB"),
for both the fermionic and bosonic components.

\begin{figure}[tbp]
\begin{center}
\includegraphics[width=.6\linewidth]{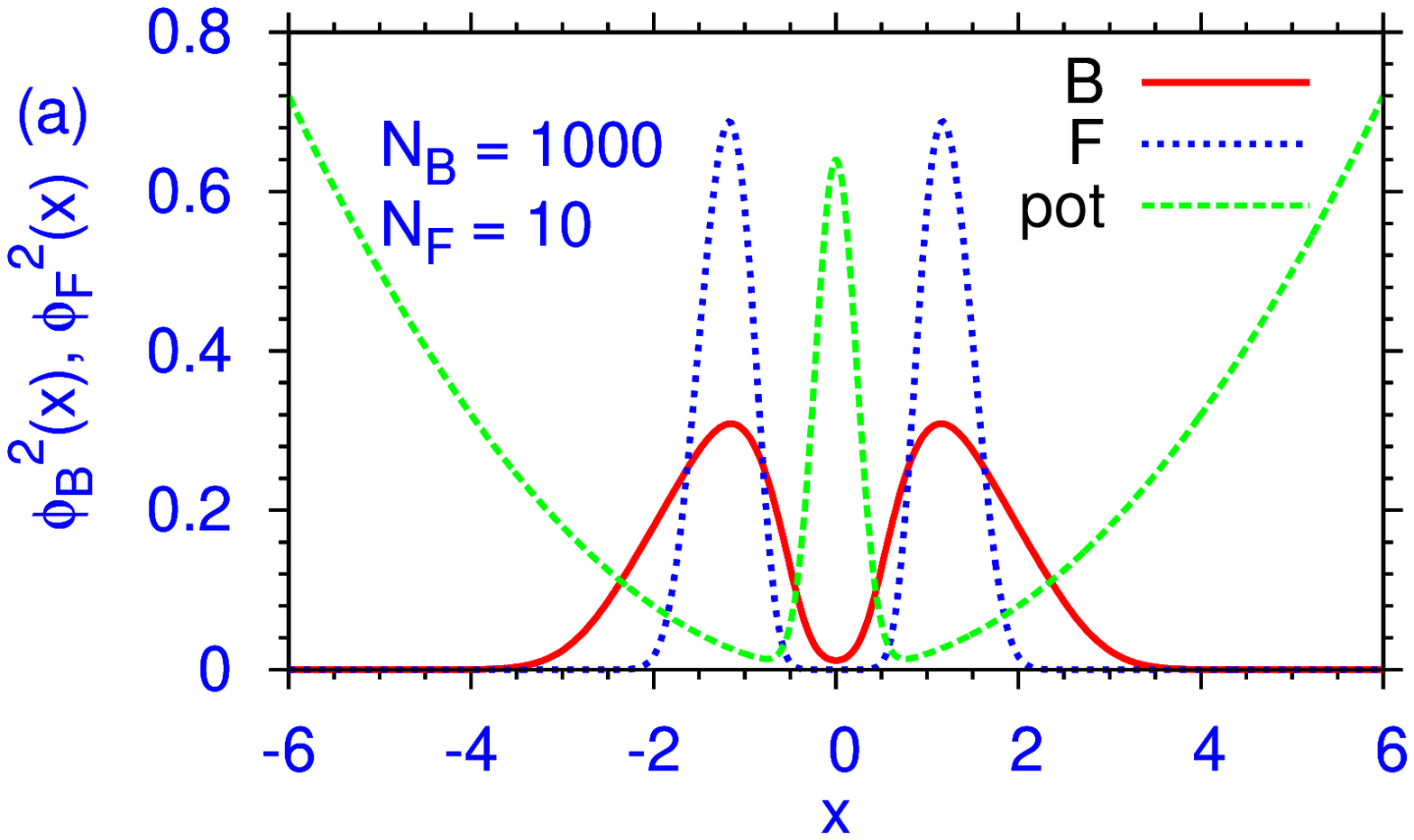} \includegraphics[width=.6%
\linewidth]{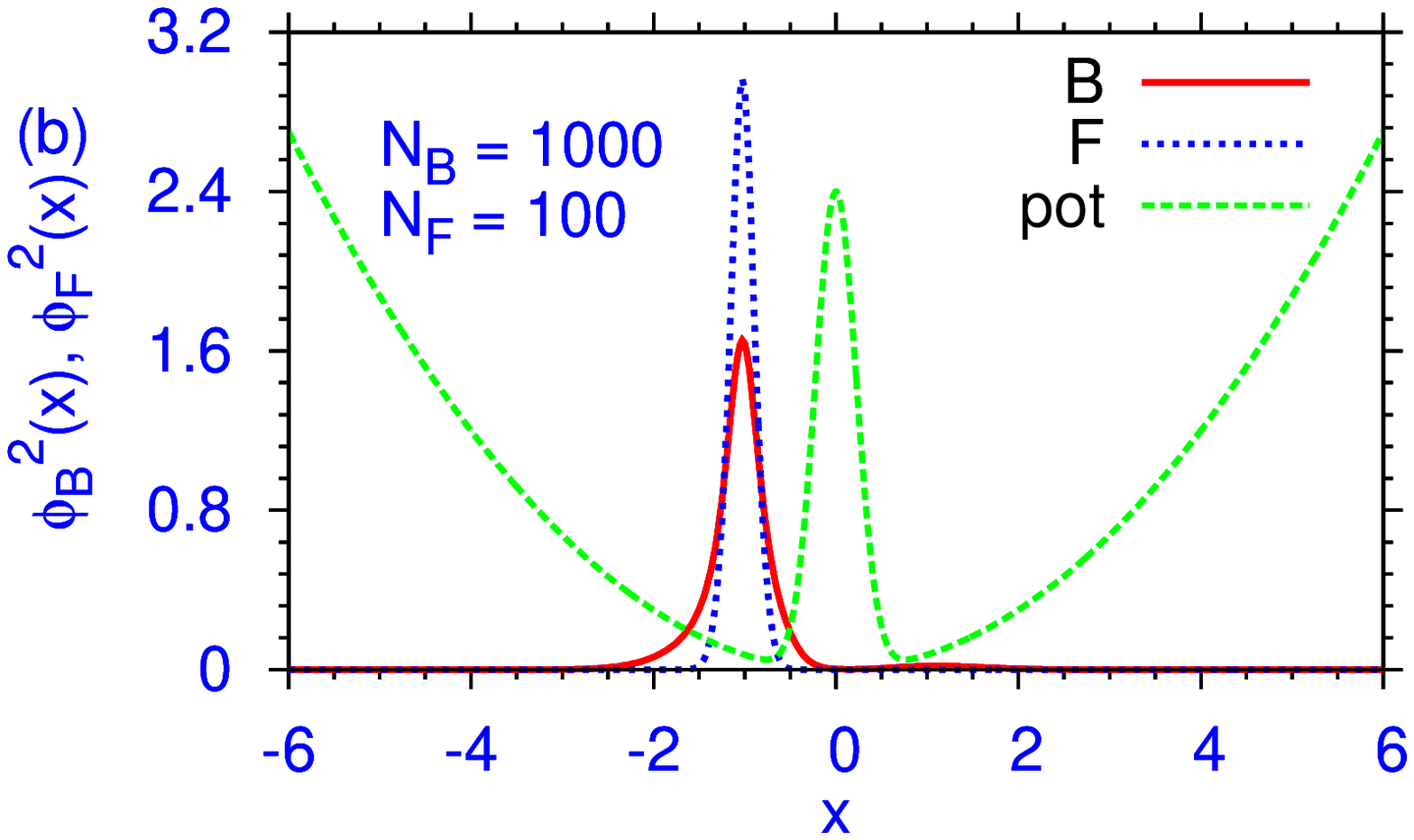}
\end{center}
\caption{(Color online). An example of the spontaneous symmetry breaking in
the $^{87}$Rb--$^{40}$K mixture trapped in potential (\protect\ref{pot-b}).
The two panels differ by the number of fermions: (a) $N_{\mathrm{F}}=10$;
(b) $N_{\mathrm{F}}=1000$.}
\label{ssb-f6}
\end{figure}

An example of the transition from the symmetric ground state of the BF
mixture to an asymmetric one, caused by the increase of the number of
fermions from $N_{\mathrm{F}}=10$ to $N_{\mathrm{F}}=1000$, while the number
of bosons trapped in potential (\ref{pot-b}) is kept constant, is displayed
in Fig. \ref{ssb-f6}. Although the applicability of the functional-density
description for $N_{\mathrm{F}}=10$ [in panel (a)] may be disputed, this
figure adequately shows the transition to the SSB.

\subsection{The case of the fermionic component in the BCS regime}

The analysis presented above pertained to the superfluid BF mixture with the
spin-balanced fermion components in the unitarity regime, where the $s$-wave
scattering length, $a_{\mathrm{F}}$, which accounts for the interaction
between the spin-up and spin-down fermions is extremely large (ideally, $a_{%
\mathrm{F}}\rightarrow \pm \infty $). Actually, essentially the same
Lagrangian (\ref{ee-f}) also applies to the BF mixture with the fermionic
component kept in the BCS regime, with $a_{\mathrm{F}}$ is negative and
small (ideally, $a_{\mathrm{F}}\rightarrow -0$). In this regime, one is
practically dealing with a gas of ideal fermions, because the respective
superfluid energy gap is exponentially small. In practical terms, to study
the system whose Fermi component falls into the BCS regime, it is sufficient
to set $\xi =1$ instead of $\xi =0.45$ in Eq. (\ref{ee-f}) \cite{manini05}.
Obviously, this change of $\xi $ will make the fermions effectively more
repulsive [see Eq. (\ref{A})], as the Pauli repulsion attains its maximum at
$\xi =1$. We have verified that, as the SSB in the mixture emerges chiefly
due to BF attraction, the increase of the intrinsic repulsion in the
Fermionic component corresponding to $\xi =1$ makes the natural BF
attraction in the $^{87}\mathrm{Rb}-^{40}\mathrm{K}$ mixture insufficient
for the appearance of the SSB for relatively small values of $N_{\mathrm{F}}$
and $N_{\mathrm{B}}$ considered above. To achieve the transition to
asymmetric ground states (for the same value of $a_{\perp }=1~\mathrm{\mu }$%
m as taken above), it is necessary to consider values of $N_{\mathrm{B}}$
and $N_{\mathrm{F}}$ exceeding $1500$ [cf. Figs. \ref{ssb-f3} and \ref%
{ssb-f5}, which display the phase diagrams of the mixture with the fermionic
component in the unitarity regime for $N_{\mathrm{B}},N_{\mathrm{F}}\leq
1000 $].

\section{The analytical approach: the Thomas-Fermi approximation for the
Bose component}

\subsection{The general case}

For the application of the analytical approach, we use the stationary
version of general equations (\ref{uB}) and (\ref{uF}), obtained by the
substitution of $\phi _{\mathrm{B,F}}(x,\tau )= \exp \left( -i\mu _{\mathrm{%
B,F}}\tau \right) u_{\mathrm{B,F}}(x)$, where chemical potentials for the
localized states must be non-positive, $\mu _{\mathrm{B,F}}\leq 0$, and real
functions $u_{\mathrm{B,F}}$ obey the following equations (with the prime
standing for $d/dx$):
\begin{equation}
-\frac{1}{2}{u}_{\mathrm{B}}^{\prime \prime }+\left[ -\mu _{\mathrm{B}}+W_{%
\mathrm{B}}(x)+g_{\mathrm{B}}u_{\mathrm{B}}^{2}+g_{\mathrm{BF}}N_{\mathrm{F}%
}u_{\mathrm{F}}^{2}\right] u_{\mathrm{B}}=0,  \label{u-B}
\end{equation}%
\begin{equation}
-\frac{1}{8}{u}_{\mathrm{F}}^{\prime \prime }+\left[ -\mu _{\mathrm{F}}+g_{%
\mathrm{F}}u_{\mathrm{F}}^{4/3}+g_{\mathrm{BF}}N_{\mathrm{B}}u_{\mathrm{B}%
}^{2}+W_{\mathrm{F}}(x)\right] u_{\mathrm{F}}=0.  \label{u-F}
\end{equation}%
An essential simplification of Eqs. (\ref{u-B}) and (\ref{u-F}) can be
achieved if the TF approximation may be applied to the former equation,
i.e., the term with the second derivative may be neglected in it \cite%
{review,rev-stringa1,lipparini}. In the present setting, with the
characteristic size of the trapped states $\Delta x\sim 1$ [see Eqs. (\ref%
{min}) and Figs. \ref{ssb-f1} and \ref{ssb-f6}], and the wave functions
subject to normalization conditions (\ref{con'}) [hence the amplitudes of
normalized density $u_{\mathrm{B}}^{2}(x)$ and $u_{\mathrm{F}}^{2}(x)$ are
also $\sim 1$], a straightforward consideration of Eq. (\ref{u-B})
demonstrates that the kinetic energy (the second derivative) is negligible
in comparison with either nonlinear term under conditions $N_{\mathrm{B}}\gg
a_{\perp }/a_{\mathrm{B}}$ or $N_{\mathrm{F}}\gg a_{\perp }/\left(
10\left\vert a_{\mathrm{BF}}\right\vert \right) $. For the value of $%
a_{\perp }=1~\mathrm{\mu }$m adopted above, and the values of $a_{\mathrm{B}%
} $ and $\left\vert a_{\mathrm{BF}}\right\vert $ for the $^{87}\mathrm{Rb}%
-^{40}\mathrm{K}$ mixture, these conditions reduce to quite realistic
inequalities, $N_{\mathrm{B}}\gg 200$, $N_{\mathrm{F}}\gg 10$.

If the TF approximation is valid, it allows one to solve Eq. (\ref{u-B}) in
the following form:
\begin{equation}
u_{\mathrm{B}}^{2}(x)=\left\{
\begin{array}{c}
g_{\mathrm{B}}^{-1}\left[ \left\vert g_{\mathrm{BF}}\right\vert N_{\mathrm{F}%
}u_{\mathrm{F}}^{2}(x)-W_{\mathrm{B}}(x)+\mu _{\mathrm{B}}\right] ,~\mathrm{%
at}~|x|<x_{0}, \\
0,~\mathrm{at}~|x|>x_{0},%
\end{array}%
\right.  \label{TF}
\end{equation}%
where it is taken into account that we are dealing with $g_{\mathrm{BF}}<0$,
and $x_{0}$ is a positive root of equation%
\begin{equation}
u_{\mathrm{F}}^{2}(x_{0})=\left( \left\vert g_{\mathrm{BF}}\right\vert N_{%
\mathrm{F}}\right) ^{-1}\left[ W_B(x_{0})-\mu _{\mathrm{B}}\right]
\label{x0}
\end{equation}
Note that the bosonic chemical potential, $\mu _{\mathrm{B}}$, is not an
arbitrary parameter; instead, it must be found from the normalization
condition (\ref{con'}), applied to expression (\ref{TF}):%
\begin{equation}
2\int_{0}^{x_{0}}\left[ \left\vert g_{\mathrm{BF}}\right\vert N_{\mathrm{F}%
}u_{\mathrm{F}}^{2}(x)-W_B(x)+\mu _{\mathrm{B}}\right] dx=g_{B}.  \label{int}
\end{equation}

\subsection{A tractable example}

The substitution of approximation (\ref{TF}) for $u_{\mathrm{B}}^{2}$ into
equation (\ref{u-F}) for the fermionic function, $u_{\mathrm{F}}(x)$, allows
one to reduce the underlying system to a single equation for $u_{\mathrm{F}%
}(x)$; however, in the general case this equation is quite complex. In
particular, the additional equation (\ref{x0}) for $x_{0}$ actually makes
the resulting equation for $u_{\mathrm{F}}$ nonlocal. Thus, in the general
case the TF approximation does not yield an explicit analytical solution.
Nevertheless, it can be obtained in a special case, when $W_{\mathrm{B}}=0$
[cf. Eq. (\ref{pot-f})] and $\mu _{\mathrm{B}}=0$. In this case, Eq. (\ref%
{TF}) yields a simple local relation,
\begin{equation}
u_{\mathrm{B}}^{2}(x)=\left( \left\vert g_{\mathrm{BF}}\right\vert N_{%
\mathrm{F}}/g_{\mathrm{B}}\right) u_{\mathrm{F}}^{2}(x),  \label{simple}
\end{equation}%
which is valid at all $x$, and Eq. (\ref{int}) reduces to a special relation
between the boson and fermion numbers,
\begin{equation}
N_{\mathrm{F}}=g_{\mathrm{B}}/\left\vert g_{\mathrm{BF}}\right\vert \equiv
-\left( a_{\mathrm{B}}/3a_{\mathrm{BF}}\right) N_{\mathrm{B}},  \label{NN}
\end{equation}%
where we have made use of Eqs. (\ref{g}); note that this approximation is
meaningful only in the case when the signs of $a_{\mathrm{B}}$ and $a_{%
\mathrm{BF}}$ are opposite. For the parameters of the $^{87}\mathrm{Rb}-^{40}%
\mathrm{K}$ mixture Eq. (\ref{NN}) amounts to $N_{\mathrm{F}}\approx 0.127N_{%
\mathrm{B}}$ . Finally, the single equation for the fermionic stationary
function takes the following form, upon the substitution of expression (\ref%
{simple}):%
\begin{gather}
-\frac{1}{8}{u}_{\mathrm{F}}^{\prime \prime }+\left[ W_{\mathrm{F}}(x)-\mu _{%
\mathrm{F}}\right] u_{\mathrm{F}}  \notag \\
+\left( 3\pi ^{2}\right) ^{2/3}\frac{3\xi }{5}N_{\mathrm{F}}^{2/3}u_{\mathrm{%
F}}^{7/3}-\frac{18a_{\mathrm{BF}}^{2}}{a_{\mathrm{B}}a_{\perp }}N_{\mathrm{F}%
}u_{\mathrm{F}}^{3}=0,  \label{F}
\end{gather}%
where we have again used Eqs. (\ref{g}). Note that the last term in Eq. (\ref%
{F}) directly illustrates the possibility proposed in this work, namely,
that the interaction mediated by the boson field may give rise to an
effective attraction in the Fermi component of the BF superfluid mixture:
indeed, the coefficient in front of this term is proportional to the square
of the scattering length, $a_{\mathrm{BF}}^{2}$, which accounts for the BF
attraction. It is also worthy to note that, in the present simplest
approximation, which makes it possible to eliminate the boson field and
reduce the model to the single equation for the fermionic function, the
attractive character of the resulting boson-mediated interaction requires $%
a_{\mathrm{B}}>0$, i.e., the \emph{repulsive} character of the direct
interaction between the bosons. Note that in the case when both the BF and
boson-boson interactions are attractive, i.e., both $a_{\mathrm{B}}$ and $a_{%
\mathrm{BF}}$ are negative, the present approximation is impossible,
according to Eq. (\ref{NN}); in the case of the BF repulsion and boson-boson
attraction, i.e., $a_{\mathrm{B}}<0$ and $a_{\mathrm{BF}}>0$, the
approximation is possible, but it leads to an effective boson-mediated
repulsion between the fermions.

Equation (\ref{F}) is a variant of the stationary NLSE with two \emph{%
competing} nonlinear terms, the self-repulsive one $\sim N_{\mathrm{F}%
}^{2/3}u_{\mathrm{F}}^{7/3}$, and the self-attractive cubic term. For $%
a_{\perp }=1~\mathrm{\mu }$m and the scattering lengths corresponding to the
$^{87}\mathrm{Rb}-^{40}\mathrm{K}$ mixture, the coefficient in front of the
cubic term is $18a_{\mathrm{BF}}^{2}/\left( a_{\mathrm{B}}a_{\perp }\right)
\approx 0.71$, while, with $\xi =0.4$ (recall it corresponds to the
unitarity regime for the fermion component), the coefficient in front of the
repulsive term is $(3\pi ^{2})^{2/3}(3\xi /5)\approx \allowbreak
2.\,\allowbreak 30$.

The SSB controlled by competing nonlinearities, \textit{viz}., self-focusing
cubic and self-defocusing quintic terms, was studied in Refs. \cite{CQ},
where it was concluded that the respective SSB diagrams, showing the
asymmetry versus the total norm of the mode (cf. Fig. \ref{ssb-f2}), tend to
form a \textit{closed loop} connecting an initial symmetry-breaking
bifurcation and a final symmetry-restoring one. The difference of Eq. (\ref%
{F}) is that here the self-focusing (cubic) term has a higher nonlinearity
power than its self-defocusing counterpart of power $7/3$, therefore no
closed-loop bifurcation diagram is expected in the present case.

The substitution of $u_{\mathrm{F}}(x)\equiv v_{\mathrm{F}}/\sqrt{N_{\mathrm{%
F}}}$ casts Eq. (\ref{F}) in a parameter-free form [the respective equation
for $v_{\mathrm{F}}$\ seems as Eq. ((\ref{F}) with $N_{\mathrm{F}}$ replaced
by $1$]; of course, this substitution changes normalization (\ref{con'}) for
the fermionic function, as the norm of $v_{\mathrm{F}}$\ \ is exactly $N_{%
\mathrm{F}}$, but not $1$. Thus, in the present approximation, the BF\
mixture is described by the universal equation. This circumstance, along
with condition (\ref{NN}) necessary for the applicability of Eq. (\ref{F}),
may explain the fact that the border of the trapped states in Fig. \ref%
{ssb-f3} is practically a straight line with the slope equal to $1$, which
runs into the SSB border at a critical value of $N_{\mathrm{F}}$ [the latter
one actually corresponds to the critical norm of field $v_{\mathrm{F}}(x)$
at which the SSB occurs in the framework of Eq. (\ref{F})].

The actual SSB\ point generated by Eq. (\ref{F}) can be predicted in an
approximate analytical form by means of the two-mode expansion, which, as
mentioned above, is commonly used for the description of SSB effects in DWP
settings \cite{two-mode}, assuming that the stationary solution is
approximated by a superposition of two linear modes, $u_{\pm }(x)$, which
are trapped in the left and right potential wells:%
\begin{equation}
u_{\mathrm{F}}(x)=A_{+}u_{+}(x)+A_{-}u_{-}(x).  \label{leftright}
\end{equation}%
Symmetric states correspond to $A_{+}=A_{-}\equiv A_{0}$ in Eq. (\ref%
{leftright}). The SSB border can be found by looking for a point where a
solution with an infinitely small antisymmetric perturbation ($\delta $), $%
A_{\pm }=A_{0}\pm \delta $, branches off from the parent symmetric state. By
performing this analysis (we do not display straightforward details here),
one finds that the value of $A_{0}$ for SSB scales inversely proportional to
$\sqrt{N_{\mathrm{F}}}$, hence the increase of the number of particles
facilitates the transition to the asymmetric ground state, as one may expect.

\section{Conclusion}

The objective of this work is to extend the analysis of the SSB
(spontaneously symmetry breaking) in DWP settings (double-well potentials),
which was recently studied in BEC and bosonic mixtures, to the BF
(Bose-Fermi) mixtures. The system is described by the GPE (Gross-Pitaevskii
equation) for the bosons, which is nonlinearly coupled to the equation for
the fermionic order parameter derived from the density functional in the
unitarity limit (in fact, a similar model also applies to the BF mixture
with the fermionic component kept in the BCS regime). Direct symmetry
breaking in the Fermi superfluid trapped in the DWP is impossible, as it
must be induced by attractive interactions, while density perturbations in
the degenerate fermionic gas interact repulsively. Nevertheless, we have
demonstrated that the SSB is possible in the mixture of $^{87}$Rb and $^{40}$%
K atoms, due to the attraction between fermions and bosons. The most
interesting case, when the effective SSB in the fermionic component could be
studied in the ``pure form", is that when the fermions are subject to the
action of the DWP, while there is no potential confining the bosons. We have
also investigated the alternative situation, with the DWP acting solely on
the bosons. Our phase diagrams in the $(N_{\mathrm{F}},N_{\mathrm{B}})$
plane, produced by means of numerical methods, clearly show that the
inter-atomic attractions can produce both symmetric and symmetry-broken
localization of the atoms which are not subject to the direct action of the
trapping potential. Applying the TF (Thomas-Fermi) approximation to the
bosonic equation, we have also developed an analytical approximation, which
allows us to reduce the model to the single equation for the fermionic
function. In the latter case, the model explicitly demonstrates the
generation of the effective attraction between fermions mediated by the
bosons.

The analysis reported in this work can be extended in other directions. A
straightforward generalization may deal with the system including the
confining potential in both components, as well as a more general analysis
of the TF approximation. A challenging possibility is to predict similar
effects in multi-dimensional Bose-Fermi mixtures.

\acknowledgments

The work of S.K.A. was partially supported by CNPq and FAPESP (Brazil).
B.A.M. acknowledges hospitality of the Department of Physics ``Galileo
Galilei" at the University of Padua, Italy, and support from the
German-Israel Foundation, through grant No. 149/2006.


\begin{thebibliography}{99}
\bibitem{bose-deg} M. H. Anderson, J. R. Ensher, M. R. Matthews, C. E.
Wieman, and E. A. Cornell, Science \textbf{269}, 198 (1995); C. C. Bradley,
C. A. Sackett, J. J. Tollett, and R. G. Hulet, Phys. Rev. Lett. \textbf{75},
1687 (1995); K. B. Davis, M.-O. Mewes, M. R. Andrews, N. J. van Druten, D.
S. Durfee, D. M. Kurn, and W. Ketterle, \textit{ibid}. \textbf{75}, 3969
(1995).

\bibitem{fermi-deg} B. DeMarco and D. Jin, Science \textbf{285}, 1703 (1999).

\bibitem{review} F. Dalfovo, S. Giorgini, L. P. Pitaevskii, and S.
Stringari, Rev. Mod. Phys. \textbf{71}, 463 (1999).

\bibitem{rev-stringa1} L. Pitaevskii and S. Stringari, \textit{Bose-Einstein
Condensation} (Clarendon Press: Oxford, 2003).

\bibitem{rev-stringa2} S. Giorgini, L. P. Pitaevskii, and S. Stringari, Rev.
Mod. Phys. \textbf{80}, 1215 (2008); H. T. C. Stoof, K. B. Gubbels, and D.
B. M. Dickrsheid, \textit{Ultracold Quantum Fields} (Springer: Dordrecht,
2009).

\bibitem{lipparini} E. Lipparini, \textit{Modern Many-Particle Physics:
Atomic Gases, Nanostructures and Quantum Liquids} (World Scientific,
Singapore, 2006).

\bibitem{manini05} N. Manini and L. Salasnich, Phys. Rev. A \textbf{71},
033625 (2005); G. Diana, N. Manini, and L. Salasnich, Phys. Rev. A \textbf{73%
}, 065601 (2006).

\bibitem{etf} L. Salasnich, N. Manini and F. Toigo, Phys. Rev. A \textbf{77}%
, 043609 (2008); F. Ancilotto, L. Salasnich, and F. Toigo, Phys. Rev. A
\textbf{79}, 033627 (2009); L. Salasnich and F. Toigo, Phys. Rev. A \textbf{%
78}, 053626 (2008); L. Salasnich, F. Ancilotto and F. Toigo, Laser Phys.
Lett. \textbf{7}, 78 (2010).

\bibitem{recent} S. K. Adhikari and L. Salasnich, Phys. Rev. A \textbf{78},
043616 (2008); S. K. Adhikari and L. Salasnich, New J. Phys. \textbf{11},
023011 (2009); S. K. Adhikari, Laser Phys. Lett. \textbf{6}, 901 (2009);
Phys. Rev. A \textbf{79}, 023611 (2009).

\bibitem{GS} S. K. Adhikari and B. A. Malomed, Phys. Rev. A 74, 053620
(2006); Europhys. Lett. \textbf{79}, 50003 (2007); Physica D \textbf{238},
1402 (2009).

\bibitem{bifurcation} G. J. Milburn, J. Corney, E. M. Wright, and D. F.
Walls, Phys. Rev. A \textbf{55}, 4318 (1997); \ A. Smerzi, S. Fantoni, S.
Giovanazzi, and S. R. Shenoy, Phys. Rev. Lett. \textbf{79}, 4950 (1997); S.
Raghavan, A. Smerzi, S. Fantoni, and S. R. Shenoy, Phys. Rev. A \textbf{59},
620 (1999); K. W. Mahmud, H. Perry, and W. P. Reinhardt, Phys. Rev. A
\textbf{71}, 023615 (2005); E. Infeld, P. Zin, J. Goca\l ek, and M.
Trippenbach, Phys. Rev. E \textbf{74}, 026610 (2006); G. Theocharis, P. G.
Kevrekidis, D. J. Frantzeskakis, and P. Schmelcher, Phys. Rev. E \textbf{74}%
, 056608 (2006); G. L. Alfimov and D. A. Zezyulin, Nonlinearity \textbf{20},
2075 (2007).

\bibitem{CQ} L. Albuch and B. A. Malomed, Mathematics and Computers in
Simulation \textbf{74}, 312 (2007); Z. Birnbaum and B. A. Malomed, Physica D
\textbf{237}, 3252 (2008).

\bibitem{dual-core} A. W. Snyder, D. J. Mitchell, L. Poladian, D. R.
Rowland, and Y. Chen, J. Opt. Soc. Am. B \textbf{8}, 2102 (1991).

\bibitem{old} E. B. Davies, Commun. Math. Physics \textbf{64}, 191 (1979).

\bibitem{Scott} J. C. Eilbeck, P. S. Lomdahl and A. C. Scott, Physica D
\textbf{16}, 318 (1985).

\bibitem{two-mode} E. A. Ostrovskaya, Y. S. Kivshar, M. Lisak, B. Hall, F.
Cattani, and D. Anderson, Phys. Rev. A \textbf{61}, 031601(2000); R.
D'Agosta, B. A. Malomed, C. Presilla, Phys. Lett. A \textbf{275}, 424
(2000); R. K. Jackson and M. I. Weinstein, J. Stat. Phys. \textbf{116}, 881
(2004); V. S. Shchesnovich, B. A. Malomed, and R. A. Kraenkel, Physica D
\textbf{188}, 213 (2004); D. Ananikian and T. Bergeman, Phys. Rev. A \textbf{%
73}, 013604 (2006); C. Wang, P. G. Kevrekidis, N. Whitaker and B. A.
Malomed, Physica D \textbf{327}, 2922 (2008); E. W. Kirr, P. G. Kevrekidis,
E. Shlizerman, and M. I. Weinstein, SIAM J. Math. Anal. \textbf{40}, 566
(2008).

\bibitem{Warsaw} A. Gubeskys and B. A. Malomed, Phys. Rev. A \textbf{75},
063602 (2007); M. Matuszewski, B. A. Malomed, and M. Trippenbach, \textit{%
ibid}. \textbf{75}, 063621 (2007); M. Trippenbach, E. Infeld, J. Goca\l ek,
M. Matuszewski, M. Oberthaler, and B. A. Malomed, \textit{ibid}. A \textbf{78%
}, 013603 (2008).

\bibitem{Heidelberg} M. Albiez, R. Gati, J. F\"{o}lling, S. Hunsmann, M.
Cristiani, and M. K. Oberthaler, Phys. Rev. Lett. \textbf{95}, 010402
(2005); for a review, see R. Gati and M. Oberthaler, J. Phys. B. \textbf{40}%
, R61 (2007).

\bibitem{skew} T. Mayteevarunyoo and B. A. Malomed, J. Opt. A: Pure Appl.
Opt. \textbf{11}, 094015 (2009).

\bibitem{4wells} C. Wang, G. Theocharis, P. G.\ Kevrekidis, N. Whitaker, K.
J. H.\ Law, D. J.\ Frantzeskakis, and B. A. Malomed, Phys. Rev. E \textbf{80}%
, 046611 (2009).

\bibitem{triple} B. Liu, L.-B. Fu, S.-P. Yang, and J. Liu, Phys. Rev. A
\textbf{75}, 033601 (2007).

\bibitem{DD} B. Xiong, J. Gong, H. Pu, W. Bao, and B. Li, Phys. Rev. A
\textbf{79}, 013626 (2009).

\bibitem{higher-order} A. Sacchetti, Phys. Rev. Lett. 1-3, 194101 (2009).

\bibitem{Dong} T. Mayteevarunyoo, B. A. Malomed, and G. Dong, Phys. Rev. A
\textbf{78}, 053601 (2008); C. Wang, P. G. Kevrekidis, N. Whitaker, D. J.
Frantzeskakis, S. Middelkamp, and P. Schmelcher, Physica D \textbf{238},
1362 (2009).

\bibitem{2comp} C. Wang, P. G. Kevrekidis, N. Whitaker and B. A. Malomed,
Physica D \textbf{327}, 2922 (2008); I. I. Satija, R. Balakrishnan, P.
Naudus, J. Heward, M. Edwards, and C. W. Clark, Phys. Rev. E \textbf{79},
033616 (2009); W. Wang, J. Phys. Soc. Jpn. \textbf{78}, 094002 (2009); C.
Lee, Phys. Rev. Lett. \textbf{102}, 070401 (2009),

\bibitem{3comp} C. Wang, P. G. Kevrekidis, N. Whitaker, T. J. Alexander, D.
J. Frantzeskakis, and P. Schmelcher, J. Phys. A Math. Theor. \textbf{42},
035201 (2009); B. Juli\'{a}-Diaz, M. Guilleumas, M. Lewenstein, A. Polls,
and A. Sanpera, Phys. Rev. A \textbf{80}, 023616 (2009); B. Juli\'{a}-Diaz,
M. Mele-Messeguer, M. Guilleumas, and A. Polls, Phys. Rev. A \textbf{80},
043622 (2009).

\bibitem{Kivshar} S. F. Caballero-Ben\'{\i}tez, E A Ostrovskaya, M. Gul\'{a}%
cs\'{\i}, and Yu. S. Kivshar, J. Phys. B \textbf{42}, 215308 (2009).

\bibitem{SKA-Pu} S. K. Adhikari, H. Lu, and H. Pu, Phys. Rev. A \textbf{80},
063607 (2009).

\bibitem{modugno} G. Modugno, G. Roati, F. Riboli, F. Ferlaino, R. J.
Brecha, and M. Inguscio, Science \textbf{297}, 2240 (2002); G. Roati, F.
Riboli, G. Modugno, and M. Inguscio, Phys. Rev. Lett. \textbf{89}, 150403
(2002).

\bibitem{sala-new} L. Salasnich, Laser Phys. \textbf{19}, 642 (2009).

\bibitem{Luca} L. Salasnich, Laser Phys. \textbf{12}, 198 (2002); L.
Salasnich, A. Parola, and L. Reatto, Phys. Rev. A \textbf{65}, 043614 (2002).

\bibitem{BFequations} K. M\o lmer, Phys. Rev. Lett. \textbf{80}, 1804
(1998); P. Capuzzi, A. Minguzzi, and M. P. Tosi, Phys. Rev. A \textbf{67},
053605 (2003); \textbf{68}, 033605 (2003); W. Yi and L.-M. Duan, Europhys.
Lett. \textbf{75}, 854 (2006); A. M. Belemuk, N. M. Chtchelkatchev, and V.
N. Ryzhov, S.-T. Chui, Phys. Rev. A \textbf{73}, 053608 (2006); Yu. V.
Bludov and V. V. Konotop, \textit{ibid}. \textbf{74}, 043616 (2006); S. K.
Adhikari, \textit{ibid}. \textbf{72,} 053608 (2005); \textit{ibid}. \textbf{%
70}, 043617 (2004); New J. Phys. \textbf{8,} 258 (2006).

\bibitem{we} S. K. Adhikari and B. A. Malomed, Phys. Rev. A \textbf{76},
043626 (2007).

\bibitem{sadhan-num} P. Muruganandam and S. K. Adhikari, Comp. Phys. Commun.
\textbf{180}, 1888 (2009).

\bibitem{mix-io} L. Salasnich and F. Toigo, Phys. Rev. A \textbf{75}, 013623
(2007); L. Salasnich, S. K. Adhikari, F. Toigo, Phys. Rev. A \textbf{75},
023616 (2007).

\bibitem{solitons} K. E. Strecker, G. B. Partridge, A. G. Truscott, and F.
G. Hulet, Nature \textbf{417}, 150 (2002); L. Khaykovich, F. Schreck, G.
Ferrari, T. Bourdel, J. Cubizolles, L. D. Carr, Y. Castin, and C. Salomon,
Science \textbf{296}, 1290 (2002).

\bibitem{Volz} T. Volz, S. D\"{u}rr, S. Ernst, A. Marte, and G. Rempe, Phys.
Rev. A \textbf{68}, 010702 (2003).
\end{thebibliography}
\end{document}